\documentclass[pra,twocolumn,showpacs,eqsecnum,superscriptaddress]{revtex4-1}

\usepackage{graphicx}
\usepackage{verbatim}
\usepackage{color}
\usepackage{amsmath}
\usepackage{microtype}
\usepackage{placeins}
\usepackage{bm}
\usepackage{amsfonts}
\usepackage{amsbsy}
\usepackage{tabularx}
\usepackage[normalem]{ulem}

\begin{document}

\newcommand{\ssum}{\sideset{}{^*}\sum}
\newcommand{\bra}[1]{\left\langle #1\right|}
\newcommand{\ket}[1]{\left|#1\right\rangle}
\newcommand{\braket}[2]{\left\langle #1|#2\right\rangle}
\newcommand{\com}[2]{\left[#1,#2\right]}
\newcommand{\braketop}[3]{\left\langle #1\left|#2\right|#3\right\rangle}
\newcommand{\mean}[1]{\left\langle #1 \right\rangle}
\newcommand{\trace}[2][]{{\rm Tr_{#1}}\left(#2\right)}
\newcommand{\ImaginaryPart}{{\rm Im}}
\newcommand{\RealPart}{{\rm Re}}
\newcommand{\leftexp}[2]{{\vphantom{#2}}^{#1}{#2}}
\newcommand{\leftind}[2]{{\vphantom{#2}}_{#1}{#2}}
\newcommand{\elem}{\in}
\newcommand{\rp}{\right)}
\newcommand{\lp}{\left(}
\newcommand{\lcb}{\left\{}
\newcommand{\rcb}{\right\}}
\newcommand{\rsb}{\right]}
\newcommand{\lsb}{\left[}
\newcommand{\lbv}{\left|}
\newcommand{\rbv}{\right|}
\newcommand{\lvb}{\lbv}
\newcommand{\rvb}{\rbv}
\newcommand{\bs}{\boldsymbol}
\renewcommand{\inf}{\infty}
\newcommand{\myfrac}[2]{^{#1\negthickspace\negthickspace}/_{\negthinspace#2}}
\newcommand{\mycaption}[2]{\caption[#1]{\small #1 #2}}
\newcommand{\order}[1]{{{\mathcal O}\lp#1\rp}}
\newcommand{\iohbar}{\frac{-i}{\hbar}}
\newcommand{\melem}[1]{_{#1}}
\newcommand{\pref}[1]{(\ref{#1})}
\renewcommand{\eqref}[1]{Eq.~\pref{#1}}
\newcommand{\hc}{\mathrm{h.c.}}

\newcommand{\qrho}{\rho}
\newcommand{\crho}{\varrho}
\newcommand{\gd}{\gamma_\downarrow}
\newcommand{\gu}{\gamma_\uparrow}

\newcommand{\varlambda}{\Lambda}
\newcommand{\varchi}{X}
\newcommand{\varS}{\mathbb{S}}
\newcommand{\varK}{\mathbb{K}}

\newcommand{\superop}[1]{{\mathcal #1}}
\newcommand{\sD}{{\superop{D}}}
\newcommand{\sL}{{\superop{L}}}
\newcommand{\sC}{{\superop{C}}}
\newcommand{\sT}{{\superop{T}}}
\newcommand{\sM}{{\superop{M}}}
\newcommand{\sB}{{\superop{B}}}

\newcommand{\tr}[1]{\mathbf{#1}}
\newcommand{\tU}{{\tr{U}}}
\newcommand{\tS}{{\tr{S}}}
\newcommand{\tD}{{\tr{D}}}
\newcommand{\tR}{{\tr{R}}}
\newcommand{\tP}{{\tr{P}}}
\newcommand{\tT}{{\tr{T}}}
\newcommand{\trans}[1]{^{#1}}

\newcommand{\atr}[1]{{\mathbf #1}}
\newcommand{\atU}{{\atr{U}}}
\newcommand{\atS}{{\atr{S}}}
\newcommand{\atD}{{\atr{D}}}
\newcommand{\atR}{{\atr{R}}}
\newcommand{\atP}{{\atr{P}}}
\newcommand{\atT}{{\atr{T}}}

\newcommand{\g}{g}
\newcommand{\e}{e}
\newcommand{\f}{f}

\newcommand{\wg}{\omega_\g}
\newcommand{\we}{\omega_\e}
\newcommand{\wf}{\omega_\f}
\renewcommand{\wr}{\omega_r}
\newcommand{\wm}{\omega_m}

\newcommand{\projector}{\Pi}
\newcommand{\proj}[1]{\projector_{#1}}
\newcommand{\dproj}[1]{\dot{\projector}_{#1}}
\newcommand{\bproj}[1]{\bar{\projector}_{#1}}

\newcommand{\ad}{a^\dag}
\newcommand{\ada}{a^\dag a}
\newcommand{\sz}{\sigma_0}
\newcommand{\szd}{\sigma_0^\dag}
\newcommand{\so}{\sigma_1}
\newcommand{\sod}{\sigma_1^\dag}
\newcommand{\gz}{g_0}
\newcommand{\go}{g_1}
\newcommand{\izp}{I_{0+}}
\newcommand{\izm}{I_{0-}}
\newcommand{\izpm}{I_{0\pm}}
\newcommand{\izmp}{I_{0\mp}}
\newcommand{\iop}{I_{1+}}
\newcommand{\iom}{I_{1-}}
\newcommand{\iopm}{I_{1\pm}}
\newcommand{\iomp}{I_{1\mp}}
\newcommand{\itp}{I_{2+}}

\newcommand{\epm}{\epsilon_m}
\newcommand{\Lz}{\Lambda_0}
\newcommand{\lz}{\lambda_0}
\newcommand{\Dz}{\Delta_0}
\newcommand{\Zz}{Z_0}
\newcommand{\chiz}{\chi_0}
\newcommand{\zetaz}{\zeta_0}
\newcommand{\xiz}{\xi_0}
\newcommand{\Lo}{\Lambda_1}
\newcommand{\lo}{\lambda_1}
\newcommand{\Do}{\Delta_1}
\newcommand{\Zo}{Z_1}
\newcommand{\chio}{\chi_1}
\newcommand{\zetao}{\zeta_1}
\newcommand{\xio}{\xi_1}

\newcommand{\subge}{\mathcal{E}_{\g\e}}
\newcommand{\sm}{\sigma_-}
\renewcommand{\sp}{\sigma_+}
\newcommand{\sigmaz}{\sigma_z}
\newcommand{\comm}[2]{\lsb #1,#2\rsb}

\newcommand{\red}{\color[rgb]{0.8,0,0}}
\newcommand{\green}{\color[rgb]{0.0,0.6,0.0}}
\newcommand{\dkgrn}{\color[rgb]{0.0,0.4,0.0}} 
\newcommand{\blu}{\color[rgb]{0,0,0.6}}
\newcommand{\blue}{\color[rgb]{0,0,0.6}}
\newcommand{\pur}{\color[rgb]{0.8,0,0.8}}
\newcommand{\blk}{\color{black}}

\title{Backaction of a driven nonlinear resonator on a superconducting qubit}
\date{\today}

\author{Maxime~Boissonneault}
 \affiliation{D\'epartement de Physique, Universit\'e de Sherbrooke, Sherbrooke, Qu\'ebec, Canada, J1K 2R1}
\author{A.~C.~Doherty}
 \affiliation{School of Mathematics and Physics, The University of Queensland, St Lucia, QLD 4072, Australia}
 \affiliation{School of Physics, The University of Sydney, Sydney, NSW 2006, Australia}
\author{F.~R.~Ong}
 \affiliation{CEA-Saclay, Gif-sur-Yvette, France}
 \affiliation{Institute for Quantum Computing and Department of Physics and Astronomy, University of Waterloo, Waterloo, Ontario, N2L 3G1, Canada}
\author{P.~Bertet}
 \affiliation{CEA-Saclay, Gif-sur-Yvette, France}
\author{D.~Vion}
 \affiliation{CEA-Saclay, Gif-sur-Yvette, France}
\author{D.~Esteve}
 \affiliation{CEA-Saclay, Gif-sur-Yvette, France}
\author{A.~Blais}
 \affiliation{D\'epartement de Physique, Universit\'e de Sherbrooke, Sherbrooke, Qu\'ebec, Canada, J1K 2R1}


\begin{abstract}
We study the backaction of a driven nonlinear resonator on a multi-level superconducting qubit. Using unitary transformations on the multi-level Jaynes-Cummings Hamiltonian and quantum optics master equation, we derive an analytical model that goes beyond linear response theory. Within the limits of validity of the model, we obtain quantitative agreement with experimental and numerical data, both in the bifurcation and in the parametric amplification regimes of the nonlinear resonator. We show in particular that the measurement-induced dephasing rate of the qubit can be rather small at high drive power. This is in contrast to measurement with a linear resonator where this rate increases with the drive power. 
Finally, we show that, for typical parameters of circuit quantum electrodynamics, correctly describing measurement-induced dephasing requires a model going beyond linear response theory, such as the one presented here.
\end{abstract}

\maketitle

\section{Introduction} 
\label{sec:introduction}
Two-level systems (TLS) and harmonic oscillators are the two simplest systems that can be described exactly with quantum mechanics. Consequently, many physical systems are described at least approximately by either of these two building blocks. As an example, in cavity quantum electrodynamics (CQED)~\cite{Haroche1992}, an atom, modeled as a TLS, interacts with a photon field inside a high quality optical or microwave resonator, modeled as a harmonic oscillator. Another example is circuit quantum electrodynamics (cQED)~\cite{Blais2004}, cavity QED's little brother and a promising candidate for the realization of a future quantum computer~\cite{Nielsen2000}. In circuit QED, a superconducting artificial atom (or qubit)~\cite{Devoret2004} is coupled to a coplanar waveguide resonator. In the context of quantum information processing, the resonator both acts as a filter, partly protecting the qubit from decoherence and relaxation, and as a measurement device for the qubit state. 

However, contrary to cavity QED where the atomic properties are fixed, the engineered devices studied in circuit QED can be tuned and are custom built. Therefore, while devices dating from the early stages of circuit QED~\cite{Blais2004,Wallraff2005} were well described by two-level systems coupled to harmonic oscillators, more recent qubits, such as the transmon~\cite{Koch2007,Schreier2008,Houck2008}, the low impedance flux qubit~\cite{Steffen2010}, and the tunable coupling qubit~\cite{Gambetta2011} are better described by multi-level systems (MLS). This is also the case for the phase qubit~\cite{Martinis2009}. Moreover, while the standard architecture for qubit readout has long been linear resonators~\cite{Blais2004}, many recent results~\cite{Siddiqi2004,Lupascu2006,Boaknin2007,Mallet2009,Ong2011} now use resonators made nonlinear with embedded Josephson junctions. Not only do these nonlinear resonators provide a bifurcation amplifier regime which considerably improves the readout --- a key requirement for quantum information processing --- but they also exhibit remarkably enriched physics. As examples, they have been used to parametrically amplify small signals~\cite{Castellanos-Beltran2007,Vijay2011} and generate squeezed light~\cite{Castellanos-Beltran2008}. 

The performance of nonlinear resonators as parametric amplifiers for small signals~\cite{Yurke2006} as well as their backaction on a qubit have also been studied theoretically~\cite{Serban2010,Laflamme2011}. However, in Refs.~\cite{Serban2010,Laflamme2011}, the qubit was assumed to be a two-level system, something which is often insufficient to understand many types of superconducting qubits. Moreover, a linear response of the output signal to the input (qubit) signal was assumed. While linear response holds away from the nonlinear resonator's critical point, where bifurcation becomes possible, and away from the switching thresholds in the bifurcation amplifier regime, we show that it breaks down close to these points. We show that linear response is unlikely to be sufficient to describe a qubit readout with a nonlinear resonator when considering typical cQED parameters. Finally, the usual dispersive theory with linear resonators assumes driving of the resonator close to its resonance frequency for measurement~\cite{Blais2004,Boissonneault2009}. As a result, the theory obtains a dependence of the ac-Stark shift on the frequency detuning between the qubit and the resonator, rather than between the qubit and the measurement drive. This is especially important when measuring with a nonlinear resonator since there is always a significant frequency detuning between the  drive and the resonator in such cases. 

In this paper, we derive a reduced qubit model going beyond these assumptions. We do so using unitary transformations, especially the dispersive~\cite{Carbonaro1979,Boissonneault2009} and the polaron transformations~\cite{Mahan2000,Irish2005,Gambetta2008}. We are especially interested in describing the ac-Stark and Lamb shifts of the qubit as well as its measurement-induced dephasing~\cite{Gambetta2006}. We note that this theory was developed in parallel to and already tested against the experimental results of Ref.~\cite{Ong2011}.

In section~\ref{sec:presentation_of_the_system}, we write the general master equation that is used to describe the multi-level qubit coupled to the nonlinear resonator. In section~\ref{sec:linear_circuit_qed_in_a_nutshell}, we recall the minimal multi-level system model of linear circuit QED in the dispersive regime. In section~\ref{sec:peculiarities_of_nonlinear_circuit_qed}, we describe the basic characteristics of nonlinear resonators and explain why we need to go beyond the assumptions given in the previous paragraphs. In section~\ref{sec:a_reduced_qubit_model}, we derive a reduced model for the qubit through a series of unitary transformations. In section~\ref{sec:backaction_on_the_qubit}, we compare the predictions of the analytical model to experimental~\cite{Ong2011} and numerical data and find quantitative agreement within the limits of the model. We also explain how the ac-Stark and Lamb shifts as well as the measurement-induced dephasing are changed by the nonlinearity of the resonator. We finally test the regime of validity of the linear response theory and show that it is unlikely to be sufficient to describe any high-fidelity qubit readout with a nonlinear resonator.

\section{Presentation of the system} 
\label{sec:presentation_of_the_system}
\begin{figure}
	\centering
	\includegraphics[width=0.95\hsize]{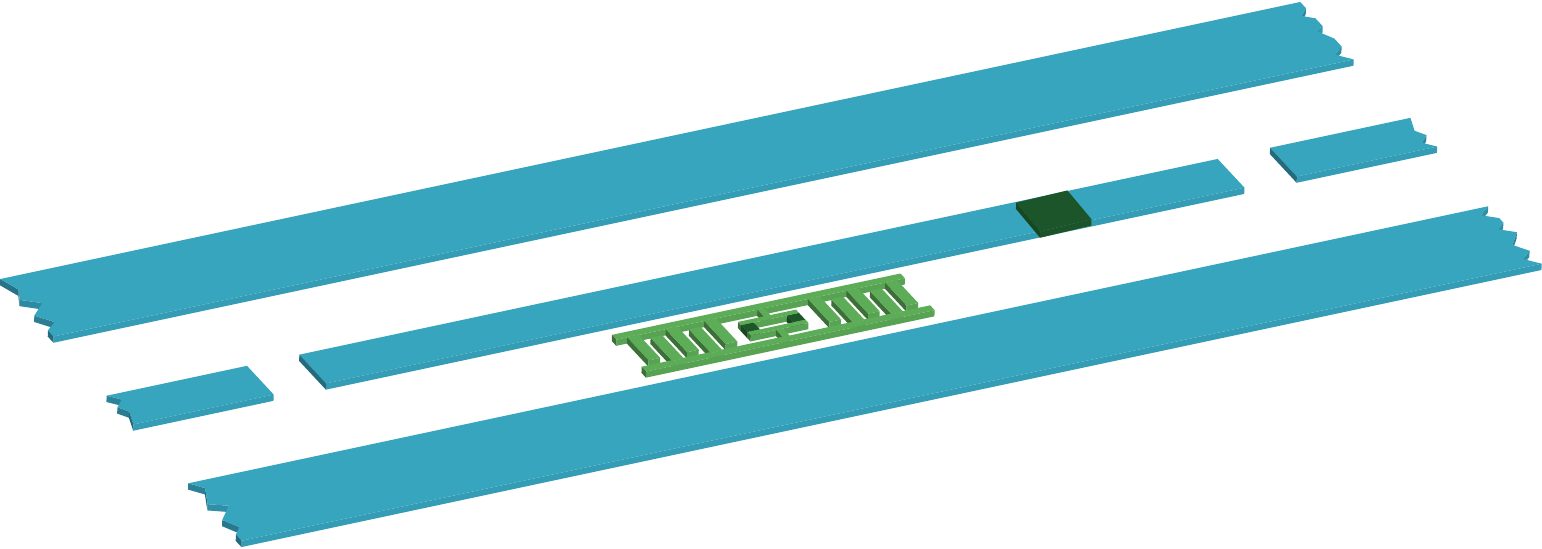}
	\caption{(Color online) Representation of one possible implementation of the system considered in this paper. This represents a stripline resonator (blue) made nonlinear with an embedded Josephson junction (dark green), capacitively coupled to a transmon qubit between the central conductor and the ground planes. The model described in this paper however applies to various other nonlinear resonators and qubits (see text). }
	\label{fig:system}
\end{figure}
We consider a system made of a multi-level qubit coupled to a nonlinear resonator. We describe the nonlinear resonator with the Hamiltonian ($\hbar=1$)~\cite{Yurke2006}
\begin{equation}
	\label{eqn:H_r}
	H_r = \wr\ada + \frac{K}{2} \ad\ad aa + \frac{K'}{3} {\ad}^3 a^3,
\end{equation}
where $a^{(\dag)}$ are the annihilation (creation) operators, $\wr$ is the resonator low-power resonance frequency, and $K$ and $K'$ are quadratic and cubic Kerr constants. Such a Kerr nonlinear resonator could be an LC-circuit with an added Josephson junction~\cite{Siddiqi2004} or a stripline resonator with one~\cite{Mallet2009} (see Fig.~\ref{fig:system}) or many~\cite{Castellanos-Beltran2007,Yamamoto2008} embedded Josephson junctions. In all these cases, the Josephson junctions act as nonlinear dissipationless inductances, rendering the resonator nonlinear. 

We describe the qubit by the generic many-level system Hamiltonian
\begin{equation}
	\label{eqn:H_q}
	H_q = \sum_{i=0}^{M-1}\omega_i\proj{i,i} \equiv \proj{\omega},
\end{equation}
where $M$ is the number of qubit levels, $\omega_i$ is the frequency of the qubit eigenstate $\ket{i}$, $\proj{i,j}\equiv\ket{i}\bra{j}$, and where we have introduced the short-handed notation
\begin{equation}
	\label{eqn:proj_x}
	\proj{x} \equiv \sum_{i=0}^{M-1} x_i \proj{i,i},
\end{equation}
which we will use on multiple occasions throughout this paper. The eigenstates $\{\ket{i}\}$ could be for example charges tunneling on and off a superconducting island such as for a Cooper-pair box~\cite{Bouchiat1998}, superposition of such charges for a transmon qubit~\cite{Koch2007} or current flowing clockwise or counterclockwise in a superconducting loop for a flux qubit~\cite{Chiorescu2003}. 

We assume a dipolar coupling between the qubit and the resonator and describe it by the interaction Hamiltonian
\begin{equation}
	\label{eqn:H_I}
	H_I = \sum_{i=0}^{M-2} g_i (\ad+a) (\proj{i,i+1} + \proj{i+1,i}),
\end{equation}
where $g_i$ are the coupling constants. The only constraint on the qubit that we impose for our model is that the selection rules only allow transitions between the qubit states $\ket{i}$ and $\ket{i\pm1}$ through the resonator. This restriction is fullfilled for good two-level qubits such as the Cooper-pair box~\cite{Bouchiat1998}, the phase~\cite{Martinis1987} and flux~\cite{Chiorescu2003} qubits, but is also realized for some more recent multi-level qubits such as the transmon~\cite{Koch2007,Schreier2008,Houck2009} and the low impedance flux qubit~\cite{Steffen2010}.

To understand the experiment of Ref.~\cite{Ong2011}, we also consider driving of the resonator. We allow for multiple qubit-detuned drives $d\in\{d_1,d_2,...,d_n\}$ as well as one spectroscopy drive $s$, quasi-resonant with the qubit frequency, that we model by the Hamiltonians
\begin{subequations}
	\begin{align}
		\label{eqn:H_d}
		H_d &= \sum_{d} \epsilon_d e^{-i\omega_d t} \ad + \epsilon_d^* e^{i\omega_d t} a, \\
		H_s	&= \epsilon_s e^{-i\omega_s t}\ad + \epsilon_s^* e^{i\omega_s t} a,
	\end{align}
\end{subequations}
where and $\epsilon_{d,s}$ and $\omega_{d,s}$ are the drives' amplitude and frequency. By \emph{quasi-resonant}, we mean that $\omega_s$ is always much closer to the $\ket{0}\leftrightarrow \ket{1}$ qubit frequency than to any other qubit transition frequencies. In experiments, these drives take the form of microwave signals sent to one port of the resonator and either transmitted to the other port or reflected back depending on the circuit design. As in the experiment of Ref.~\cite{Ong2011}, we will later on take the amplitude of the spectroscopy drive $\epsilon_s$ to be small such that its contribution to the intra-resonator field is small. The case of high amplitude spectroscopy will be treated in a following publication~\cite{Boissonneault2011b}.

Finally, to model dissipation, we use the Lindblad-type master equation
\begin{equation}
	\begin{split}
		\label{eqn:master_equation}
		\dot\rho &= -i\comm{H}{\rho} + \kappa\sD[a]\rho + \kappa_{\rm NL}\sD[a^2]\rho \\
		&\quad + \gamma \sum_{i=0}^{M-2} \lp\frac{g_i}{g_0}\rp^2 \sD\lsb \proj{i,i+1} \rsb\rho + 2\gamma_\varphi \sD\lsb \proj{\varepsilon} \rsb\rho,
	\end{split}
\end{equation}
where
\begin{equation}
	\label{eqn:dissipator}
	\sD[A]\rho \equiv \frac{1}{2} (2A\rho A^\dag - A^\dag A\rho - \rho A^\dag A),
\end{equation}
and $H = H_r+H_q+H_I+H_d+H_s$. In this master equation, $\kappa$ and $\kappa_{\rm NL}$ are the resonator's rates of one- and two-photon loss~\cite{Yurke2006}, $\gamma$ is the qubit $\ket{1}\rightarrow \ket{0}$ decay rate and $\gamma_\varphi$ is the qubit pure dephasing rate for the same states. For $\proj{\varepsilon}$, we defined $\varepsilon_i \equiv \frac{\partial (\omega_i - \omega_0)}{\partial X} \times \lp \frac{\partial(\omega_1-\omega_0)}{\partial X}\rp^{-1}$ as the $X$-dispersion, where $X$ is some control parameter (could be flux or charge for example), with $\varepsilon_0=0$ and $\varepsilon_1 = 1$ by definition.

This master equation can be obtained by modeling the coupling of the qubit and the resonator to baths of harmonic oscillators and then tracing over the baths~\cite{Carmichael2002}. When obtaining this master equation, we made three assumptions. First, we assumed that the noise spectra are white around the relevant frequencies for relaxation ($\sim$~GHz) and dephasing ($<1$~MHz). For this approximation to hold, the baths must be white on a frequency range comparable to the resonator or qubit linewidths. While this approximation should hold for relaxation ($\sim$~GHz frequencies) if the resonator and the qubit have high quality factors, it may fail for dephasing ($<1$~MHz frequencies) if, for example, the noise has a $1/f$ spectrum and hence varies by many orders of magnitudes over a single resonator or qubit linewidth. In this latter case, one needs to be more careful and take the noise spectrum into account when deriving the master equation~\cite{Ithier2005,Koch2007}. Second, we assumed that the noise causing qubit relaxation couples to the qubit through dipolar interaction, yielding the scaling in $g_i/g_0$ for the $\gamma$ dissipator. Finally, we considered that dephasing is caused by (white) noise at low frequencies in the control parameter $X$.

\section{Linear circuit QED in a nutshell} 
\label{sec:linear_circuit_qed_in_a_nutshell}
Before going to the nonlinear case, it is useful to review some aspects of the more standard linear case. In linear circuit QED, one is interested in the system described in section~\ref{sec:presentation_of_the_system}, but with $K=K'=\kappa_{\rm NL}=0$, and with a qubit which can have two or more states. Many aspects of this system have been studied extensively both theoretically and experimentally, ranging from qubit measurement~\cite{Blais2004,Bianchetti2009} and single- and two-qubit gates~\cite{Wallraff2005,Rigetti2005,Blais2007,Majer2007,Haack2010,Wu2010} to dissipation and dephasing~\cite{Schuster2005,Gambetta2006,Gambetta2008,Houck2008,Boissonneault2008,Boissonneault2009,Wilson2010}. In this section, we present the minimal theory of the dispersive regime where the couplings $g_i$ are much smaller than the qubit-resonator detunings $\Delta_{i,j} \equiv \omega_{ij} - \omega_r \equiv \omega_{i} - \omega_{j} - \omega_r$. In this regime, there is no direct exchange of energy between the qubit and the resonator, and most of the physics can be understood from an approximate diagonalization of the undriven Hamiltonian $H_r+H_q+H_I$~\cite{Blais2004}. To second order in perturbation theory and assuming that the qubit is a TLS, this diagonalization yields
\begin{equation}
	\label{eqn:dispersive_hamiltonian}
	H_{\rm disp} = \frac{\omega_{10} + \chi}{2}\sigma_z + \omega_r\ada + \chi\sigma_z\ada,
\end{equation}
where the effective qubit frequency is Lamb-shifted by a quantity $\chi=g^2/\Delta_{1,0}$. The last term of this Hamiltonian can either be seen as a qubit state-dependent pull of the resonator frequency --- which allows for qubit measurement~\cite{Blais2004} --- or as an ac-Stark shift of the qubit frequency that depends on the number of photons in the resonator~\cite{Schuster2005}.

\begin{figure}
	\centering
	\includegraphics[width=0.95\hsize]{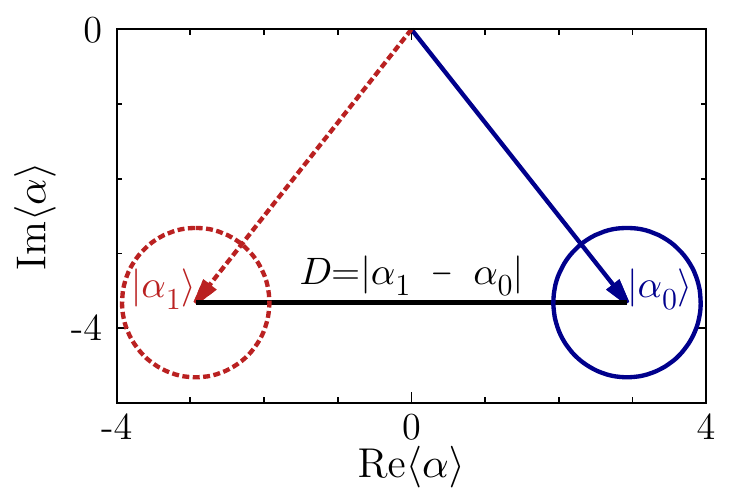}
	\caption{(Color online) Phase space representation of the pointer states $\alpha_0$ (full blue circle) and $\alpha_1$ (red dashed circle) given by \eqref{eqn:alpha_dispersive}. Parameters are $\chi/2\pi=4~$MHz, $\kappa/2\pi=10$~MHz, $\omega_p=\omega_r$ and $\epsilon_p/2\pi=30$~MHz. }
	\label{fig:pointers}
\end{figure}

In addition to the Lamb and ac-Stark shifts of the qubit frequency, the qubit's coupling to the driven resonator leads to additional sources of relaxation and dephasing. Among these are Purcell relaxation~\cite{Houck2008} in which the qubit relaxes through the resonator's photon loss channel, dressed dephasing~\cite{Boissonneault2008,Boissonneault2009,Wilson2010} in which pure dephasing of the dressed qubit-resonator states leads to effective relaxation and heating of the qubit, and measurement-induced dephasing~\cite{Schuster2005,Gambetta2006,Gambetta2008} which is the unavoidable dephasing caused by acquisition of information about a quantum system. For a linear resonator and in a dispersive measurement regime, it is shown in Refs.~\cite{Gambetta2006,Gambetta2008} that the measurement-induced dephasing rate is given by
\begin{equation}
	\Gamma_{\varphi m} = \frac{\kappa D^2}{2} \propto \frac{\kappa}{2} \bar{n},
\end{equation}
where $D = |\alpha_1 - \alpha_0|$ is the distinguishability of two pointer states of the resonator and $\bar{n}$ is the average number of photons inside the resonator. Under resonator driving, the pointer state $\alpha_{i}$ is the coherent state $\ket{\alpha_i}$ that represents the resonator's field if the qubit is in the state $\ket{i}$. For a linear resonator and a two-level system described by the dispersive Hamiltonian Eq.~(\ref{eqn:dispersive_hamiltonian}) with a single added drive of amplitude $\epsilon_p$ and frequency $\omega_p$, these coherent states are given by
\begin{equation}
	\label{eqn:alpha_dispersive}
	\alpha_{1/0} = \frac{i\epsilon_p}{-i(\omega_r-\omega_p\pm\chi) - \kappa/2},
\end{equation}
and are represented in phase space on Fig.~\ref{fig:pointers} for a resonant drive.

The distance $D$ between these pointer states in phase space depends on the cavity pull $\chi$ and, for a dispersive measurement with a linear resonator, increases with the number of photons or equivalently with the strength of the measurement drive. It is further shown in Ref.~\cite{Gambetta2008} that, in the linear case, the measurement-induced dephasing rate reaches the smallest value permitted by quantum mechanics. In other words, it saturates the inequality
\begin{equation}
	\label{eqn:quantum_limit}
	\Gamma_{\varphi m} \ge \frac{\Gamma_{\rm meas.}}{2},
\end{equation}
where $\Gamma_{\rm meas.}$ is the measurement rate~\cite{Clerk2003}, corresponding to the rate at which information is gained on the system being measured. One of the questions that we will try to answer in this paper is whether or not this inequality can be saturated when using a nonlinear resonator for homodyne dispersive measurement of the qubit.  

\section{Features specific to nonlinear circuit QED} 
\label{sec:peculiarities_of_nonlinear_circuit_qed}
\begin{figure}
	\centering
	\includegraphics[width=0.95\hsize]{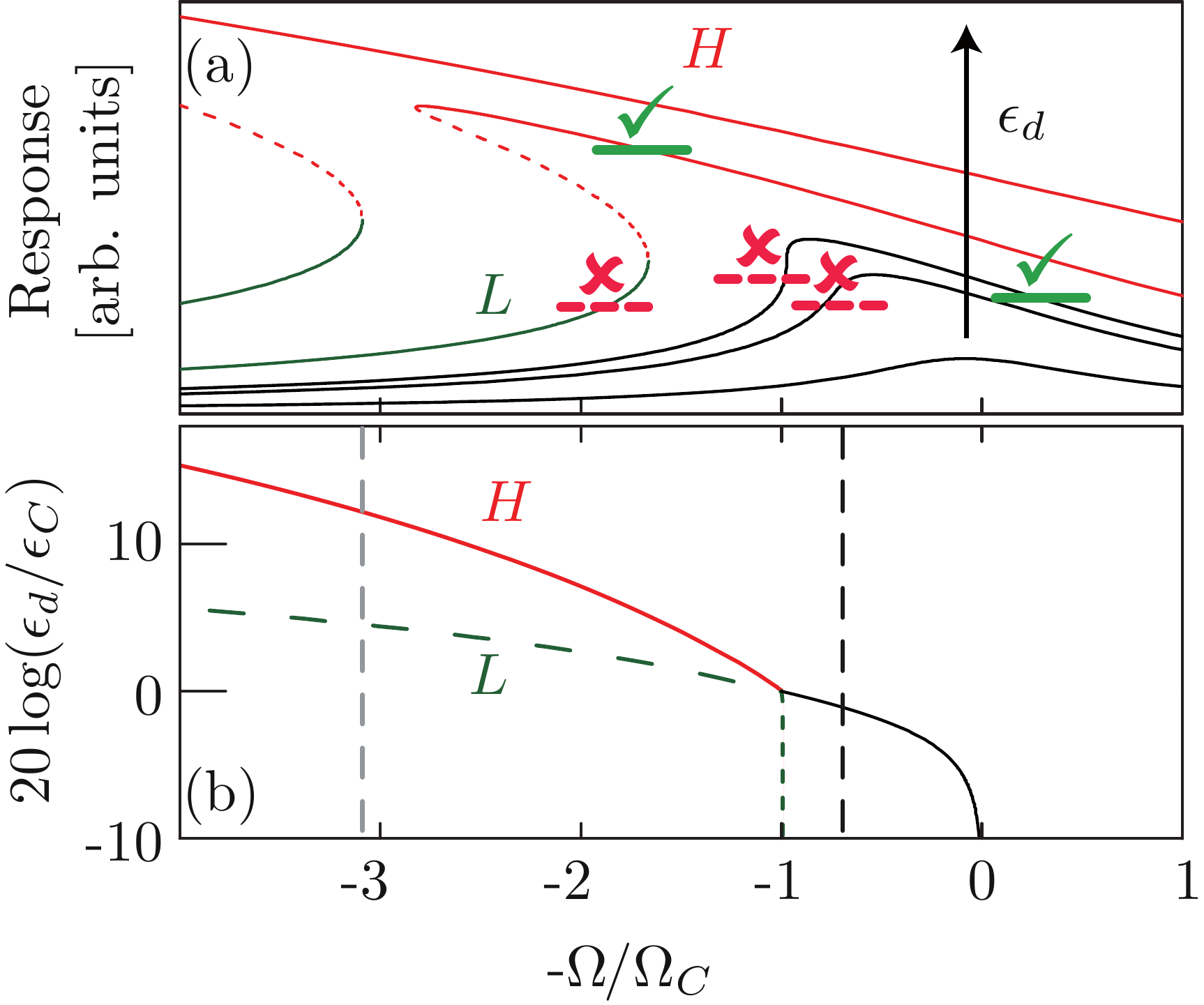}
	\caption{(Color online) (a) Amplitude of the resonator internal field (arbitrary units) in response to a drive of reduced frequency $\Omega=2(\omega_r-\omega_d)/\kappa$ ($\Omega_C=\sqrt{3}$) for increasing drive amplitudes $\epsilon_d$. The horizontal lines indicate regions for which linear response theory would (full green lines, check marks) and would not (dashed red lines, X marks) be valid for modeling a qubit-resonator system. (b) Stability diagram of the resonator (see text for explanations). As discussed in subsection~\ref{sub:experiment_and_qubit_spectra}, the vertical dashed lines represent the two operating points studied in this paper.}
	\label{fig:bifurcation_phase_diagram}
\end{figure}

Depending on the amplitude $\epsilon_d$ and frequency $\omega_d$ of the drive, the response of a Kerr nonlinear resonator can be either mono- or bi-valuated. The stability diagram describing this behavior can be parametrized by the reduced detuning frequency $\Omega \equiv 2(\omega_r-\omega_d)/\kappa$ and by the drive amplitude $\epsilon_d$. If the reduced detuning is smaller than --- but close to --- a critical value $\Omega_C=\sqrt{3}$, the nonlinear resonator can be used as a low-noise parametric amplifier~\cite{Castellanos-Beltran2007}. This has been used recently to amplify microwave signals at the single photon level~\cite{Vijay2011}. For $\Omega/\Omega_C>1$, the stability diagram, illustrated in Fig.~\ref{fig:bifurcation_phase_diagram}(b), shows two bistability thresholds~\cite{Siddiqi2005}. Below the first one (dashed green line), a low ($L$) amplitude response [see Fig.~\ref{fig:bifurcation_phase_diagram}(a)] of the resonator is observed. Above the second one (full red line), one rather observes a high ($H$) amplitude response. Between the two thresholds, both $L$ and $H$ are stable. Because of the coupling to the qubit, this stability diagram depends on the qubit state. This dependence allows the nonlinear resonators to be used as a sample-and-hold detector as has been demonstrated in Refs.~\cite{Siddiqi2006,Lupascu2006,Mallet2009}. 

Before going forward with the theory, we want to highlight two peculiarities of circuit QED with a nonlinear resonator that are often overlooked. These two aspects --- the detuning of the readout drive from the resonator frequency and the limits of the linear response theory --- as well as their impact on the theory are discussed further in the following subsections. 

\subsection{Detuned measurement drive} 
\label{sub:detuned_measurement_drive}
Both in usual low power dispersive measurement of a TLS~\cite{Blais2004} and the more recent high power avalanche readout~\cite{Reed2010,Boissonneault2010,Bishop2010}, measurement with a linear resonator is done with a drive at or very close to the resonator frequency $\omega_r$. On the contrary, measurement with a nonlinear resonator is always done with a drive source significantly detuned from $\omega_r$~\cite{Siddiqi2006a,Siddiqi2006,Lupascu2006,Mallet2009}. As can be seen from in Fig.~\ref{fig:bifurcation_phase_diagram}, this detuning is required to bias the system either in the region of highest parametric gain or in the bistability region. 

Because of the Jaynes-Cummings interaction, the drive on the resonator also acts on the qubit. Since the cavity is acting as a filter, the effective drive amplitude as seen by the qubit is expected to scale as $1/(\omega_r-\omega_d)$. Photons entering the cavity because of this drive will cause an ac-Stark shift of the qubit $\chi \mean{\ada}$. The shift per photon $\chi$ should depend on the drive frequency. This is however not the case for the usual expressions for a TLS, where $\chi=\chi_{0}$~\cite{Blais2004},  or for a MLS, where $\chi=2\chi_0 - \chi_1$~\cite{Koch2007}, with $\chi_i \equiv g_i^2/\Delta_{i+1,i} = g_i^2/(\omega_{i+1,i}-\omega_r)$. Indeed, these expressions scale with the inverse of the qubit-resonator detuning. One would rather expect to find $\chi_i \equiv g_i^2/(\omega_{i+1,i}-\omega_d)$ since the drive photons are at frequency $\omega_d$. While a relative change of a few percents on $\Delta_{1,0}$ yields the same relative change on $\chi$ for a the two-level system, the effect can be twice as big for a MLS because of the reduced value of $\chi$. To obtain quantitative agreement with the results of Ref.~\cite{Ong2011}, we obtain below an expression for the ac-Stark that contains the expected frequencies.

\subsection{Limits to the validity of linear response in circuit QED} 
\label{sub:limits_to_the_validity_of_linear_response_in_circuit_qed}
As stated before, Kerr oscillators have been used experimentally as parametric amplifiers for small signals. They have also been studied theoretically extensively. As examples, Yurke and Buks have studied their performance and calculated their gain~\cite{Yurke2006}, while Laflamme and Clerk have shown that these amplifiers are not quantum limited in the sense of \eqref{eqn:quantum_limit} for a qubit measurement~\cite{Laflamme2011}. Moreover, these last authors show that the quantum limit can be reached if one makes use of correlations between the resonator and the system coupled to it. 

These two results were however obtained in the limit of linear response theory. In this limit, one finds the driven resonator's stationary state $\bar\alpha$ without the coupling to the qubit and then expands the solution including the qubit around the stationary solution $\alpha \approx \bar\alpha + \delta\alpha$. For a qubit measurement, the signal that is amplified by the resonator takes the form of a pull $\pm\chi$ of the resonator frequency which in turns depend on the qubit state as expressed in \eqref{eqn:dispersive_hamiltonian}. For a linear resonator in the dispersive regime, the $\alpha_i$'s given by \eqref{eqn:alpha_dispersive} can be rewritten as
\begin{equation}
	\label{eqn:alpha_dispersive_linear_response}
	\begin{split}
		\alpha_{1/0} &= \bar\alpha\lp 1 + \frac{\pm i\chi}{-i(\omega_r-\omega_p \pm \chi) - \kappa/2}\rp, \\
		&\approx \bar\alpha\lp 1 + \frac{\pm i\chi}{-i(\omega_r-\omega_p) - \kappa/2}\rp,
	\end{split}
\end{equation}
where the linear response expressed by the second line holds if $|i(\omega_r-\omega_p)+\kappa/2| \gg |\chi|$. Therefore, the validity of linear response in this linear dispersive case is not affected by the driving strength, but is rather determined by the ratio $2\chi/\kappa \ll 1$. 

This analysis however does not hold for a nonlinear resonator. Indeed, in order for linear response theory to stay valid with a nonlinear resonator, $\alpha$ must change linearly with the pulled frequency --- or equivalently with the drive-resonator detuning --- over a frequency range $2\chi$. While for a linear resonator, it has been shown~\cite{Gambetta2006} that the optimal SNR is obtained for $2\chi=\kappa$, the improved measurement efficiency  with a nonlinear resonator allows for smaller cavity pulls. Taking $\chi=0.2\kappa$ as a typical value of the cavity pull translates into a range of $\Omega/\Omega_C \sim 0.5$ over which the signal must be linear in frequency for the linear response to stay valid. This range is illustrated on Fig.~\ref{fig:bifurcation_phase_diagram}(a) with the horizontal lines. The full green lines represent regimes for which linear response would be a good approximation, while dashed red lines represent regimes for which the response is not linear over the appropriate range. We argue that the linear response approximation will break down as soon as the slope of the response --- and hence the gain of the amplifier --- becomes significant.

In the following section, we derive a theory that goes further than linear response theory using the polaron transformation approach of Ref.~\cite{Gambetta2008}. 


\section{Reduced qubit model} 
\label{sec:a_reduced_qubit_model}
In this section, we derive a reduced qubit model that captures the ac-Stark and Lamb shift of the qubit transition frequencies as well as  measurement-induced dephasing. This is done by performing unitary transformations on the master equation~(\ref{eqn:master_equation}). These transformations have two objectives. First, transforming the system into its eigenbasis in which the physics is easier to understand. Second, eliminating the resonator to obtain a master equation for the qubit alone. 

In order to reach these objectives, many transformations have been used in the litterature. The dispersive transformation~\cite{Carbonaro1979,Boissonneault2008} (here generalized for a MLS)
\begin{equation}
	\label{eqn:quantum_dispersive_transformation}
	\tD = \exp\lsb \sum_{i=0}^{M-2} \lambda_i \ad \proj{i,i+1} - \lambda_i^* a\proj{i+1,i}\rsb,
\end{equation}
where $\lambda_i=g_i/(\omega_{i+1,i}-\omega_r)$, diagonalizes the Jaynes-Cummings Hamiltonian and reveals the Lamb and ac-Stark shifts. This transformation however only knows about photons that are at the resonator frequency $\omega_r$ and fails to correctly model the measurement drive-resonator frequency detuning as discussed in Sec.~\ref{sub:detuned_measurement_drive}. 

Another useful transformation is the displacement operator~\cite{Scully1997},
\begin{equation}
	\label{eqn:displacement_transformation}
	D(\alpha) = \exp\lsb \alpha\ad - \alpha^* a\rsb,
\end{equation}
which displaces a coherent state $\ket{-\alpha}$ of a resonator to the ground state $\ket{0}$. In operator representation, it corresponds to the change $a\rightarrow a+\alpha$, where $\alpha$ represents the classical average field and $a$ its quantum fluctuations. Doing this transformation before the dispersive transformation, as was done for example in Ref.~\cite{Blais2007}, yields the correct qubit-drive detuning in the ac-Stark shift. The ac-Stark shift then depends on the mean field amplitude $\alpha$ and the ac-Stark shift per photon depends on the drive-qubit frequency. However, doing this transformation in the context of a nonlinear resonator is akin to doing a linear response theory. Indeed, it is the same as assuming that the intra-resonator field is $\ket{\alpha}$ and then look at all further perturbation, such as the cavity-pull, with respect to this mean field value. This will be discussed further in Sec.~\ref{sub:measurement_induced_dephasing}.

A third transformation that was used in Ref.~\cite{Gambetta2008} to calculate the measurement-induced dephasing rate, as well as in Ref.~\cite{Irish2005} to study a qubit coupled to a mechanical resonator beyond the rotating wave approximation (RWA), is the polaron transformation~\cite{Mahan2000} (here generalized for a MLS)
\begin{equation}
	\label{eqn:polaron_transformation}
	\tP = \sum_{i=0}^{M-1} \proj{i,i} D(\alpha_i).
\end{equation}
This corresponds to a displacement transformation that is conditional on the qubit state. It allows for different cavity states $\ket{\alpha_i}$ for each qubit state $\ket{i}$, which makes it possible to go beyond the linear response approximation. It is important to note that the field amplitudes $\alpha_i$ are free parameters in this transformation. In practice, these amplitudes will be chosen such as to cancel specific terms in the transformed Hamiltonian. Moreover, and as will become clear below, these different $\alpha_i$'s will be independent solutions of qubit-state-dependent nonlinear equations, and not expansions around a mean solution $\bar\alpha$ of a single mean nonlinear equation. 

In the following subsections, we perform three transformations in order to approximately diagonalize the Hamiltonian and transform the full master equation~\eqref{eqn:master_equation} to a reduced qubit master equation containing all the relevant physics needed to account the low power spectroscopy of a qubit coupled to a nonlinear resonator  driven by an external field.

\subsection{Polaron frame} 
\label{sub:polaron_frame}
While the polaron transformation can be performed exactly on terms that are diagonal in the qubit subspace, applying it on non-diagonal terms unfortunately yields complicated expressions. For example, applying it on a qubit ladder operator $\sigma_-$ yields
\begin{equation}
	\tP^\dag \sigma_- \tP = \sigma_- D(\alpha_1 - \alpha_0) e^{-i\ImaginaryPart[\alpha_1^*\alpha_0]},
\end{equation}
which, through the displacement operator $D$, contains all powers of $a$ and $\ad$. For this reason, the polaron transformation was used in Refs.~\cite{Gambetta2008,Boissonneault2008,Boissonneault2009} after doing the dispersive transformation which eliminates the off-diagonal qubit operators. In this paper, we instead apply it before the dispersive transformation, assume that $|\alpha_1-\alpha_0|\ll1$ and take as a simplification $\tP^\dag\sigma_-\tP \approx \sigma_-$. The small distinguishability approximation $|\alpha_1-\alpha_0|<1$ will be made throughout this calculation and will limit the range of validity of the theory in a way which will be discussed later.

The application of the polaron transformation on the master equation~(\ref{eqn:master_equation}) is presented in Appendix~\ref{annsec:polaron_transformation}. Following this Appendix, we use the notation $H_i'$ to represent a part of the Hamiltonian in this first transformed frame that contains $i$ resonator ladder operators $a^{(\dag)}$. First, for $i=0$ corresponding to the qubit-only Hamiltonian we find
\begin{equation}
	\begin{split}
		\label{eqn:H_0_prime}
		H_0' &= \proj{\omega} + \sum_{i=0}^{M-2} g_i\lsb \proj\alpha^*\proj{i,i+1}+\proj{i+1,i}\proj\alpha\rsb \\
		&\quad + \omega_r |\proj\alpha|^2 - \ImaginaryPart[\proj\alpha\dproj\alpha^*] + \frac{K}{2}|\proj\alpha|^4 + \frac{K'}{3}|\proj\alpha|^6 \\
		&\quad + \sum_{d}\lsb \epsilon_d e^{-i\omega_d t}\proj\alpha^* + \mathrm{h.c.}\rsb, \raisetag{16pt}
	\end{split}
\end{equation}
where $\proj{\alpha}$ is defined according to \eqref{eqn:proj_x}. In this Hamiltonian, the second term of the first line acts as drives on the qubit at the frequencies contained in the time dependence of $\alpha$. The last two lines will be partly cancelled below by the choice of $\alpha$ given in \eqref{eqn:condition_alpha_d} and we will neglect the small remaining parts. 

We also obtain the qubit-resonator Hamiltonian, limited to terms with one resonator ladder operator,
\begin{equation}
	\label{eqn:H_1_prime}
	H_1' = G' \ad + {G'}^\dag a + \sum_{i=0}^{M-2} g_i(\ad+a)(\proj{i,i+1}+\proj{i+1,i}),
\end{equation}
where
\begin{equation}
	\begin{split}
		G' &\equiv (\omega_r-i\tfrac{\kappa}{2})\proj\alpha + (K-i\kappa_{\rm NL})|\proj\alpha|^2\proj\alpha \\
		&\quad + K'|\proj\alpha|^4\proj\alpha + \sum_{d} \epsilon_d e^{-i\omega_d t} -i\dproj\alpha. \\
	\end{split}
\end{equation}
We will see below that the two first terms of $H_1'$ can be cancelled by a proper choice of $\proj{\alpha}$. The last term will yield the Lamb shift of the qubit frequencies once the dispersive transformation is done. 

Finally, we find for the Hamiltonian containing terms with two resonator ladder operators
\begin{equation}
	\label{eqn:H_2_prime}
	H_2' = \omega_r'(\alpha)\ada + \lp\Upsilon{\ad}^2+\Upsilon^*a^2\rp, 
\end{equation}
where
\begin{subequations}
	\begin{align}
		\omega_r'(\alpha) &\equiv \omega_r+2K|\proj\alpha|^2+3K'|\proj\alpha|^4, \\
		\Upsilon &\equiv \lp\tfrac{K-i\kappa_{\rm NL}}{2} + K'|\proj\alpha|^2\rp\proj\alpha^2.
	\end{align}
\end{subequations}
With $|\proj{\alpha}|^2 = \sum_{i=0}^{M-1} |\alpha_i|^2 \Pi_{i,i}$ corresponding to the number of photons associated to the different qubit states, we see from the expression for $\omega_r'(\alpha)$ that the resonator frequency is changed by the nonlinearity as expected. Moreover, the last term of \eqref{eqn:H_2_prime} will squeeze the resonator field. This will be studied in elsewhere~\cite{Boissonneault2011b} and, for the scope of this paper, we will consider squeezing to be negligible.

Having transformed the Hamiltonian, we now apply the polaron transformation to the dissipative parts of the master equation~\eqref{eqn:master_equation}. We note that one could alternatively apply the transformations on the system-bath Hamiltonians before deriving the master equation. In this way, it would be possible to relax the white noise approximation~\cite{Boissonneault2009}, something we will not focus on here. Applying the transformation, we arrive at the master equation of the system in the polaron frame
\begin{equation}
	\begin{split}
		\label{eqn:master_equation_prime}
		\dot\rho' &= -i\comm{H_0'+H_1'+H_2'}{\rho'} + \kappa\sD[a]\rho' + \kappa_{\rm NL}\sD[a^2]\rho' \\
		&\quad + \gamma \sum_{i=0}^{M-2} \lp\frac{g_i}{g_0}\rp^2 \sD\lsb \proj{i,i+1} \rsb\rho' + 2\gamma_\varphi \sD\lsb \proj{\varepsilon} \rsb\rho' \\
		&\quad + \kappa\sD[\proj{\alpha}]\rho' + \kappa_{\rm NL}\sD[\proj{\alpha}^2]\rho' + 4\kappa_{\rm NL}\sD[a\proj{\alpha}]\rho'.
	\end{split}
\end{equation}
When obtaining the dissipative terms, we have neglected non-Linbladian terms of the form $a[\rho',\proj{\alpha}^*]$ under the assumption that in the polaron frame, the resonator is in, or close to, its ground state (see Appendix~\ref{annsec:polaron_transformation} and Ref.~\cite{Gambetta2008}). In this equation, the two first lines are the Hamiltonian part as well as the unchanged parts of the dissipative terms. The last line contains measurement-induced dephasing through the single-photon (first term) and two-photon (second term) loss decay channel, as well as some additional resonator decay (last term).

In this polaron frame, we end up with a resonator whose frequency is shifted by the nonlinearity and the amplitude of the classical fields $\alpha_i$. This resonator is driven with an adjustable strength $G$ which could be set to zero by a proper choice of $\alpha_i$. It is important to note that we did not make that choice yet because, if we did, we would have $\alpha_i=\alpha_j$ and therefore would lose all dependence of the field amplitudes $\alpha_i$ over the qubit state. The choice of the value of the qubit-state dependent fields $\alpha_i$ will be made only after moving, in the next subsection, to what we call the classical dispersive frame. Finally, in the polaron frame, the qubit is driven off-resonantly at frequencies $\omega_d$ and quasi-resonantly at frequency $\omega_s$ with amplitudes $\alpha_{i,d}$  and $\alpha_{i,s}$. As we will now show, the off-resonant drives will yield the correct ac-Stark shifts of the qubit frequencies. 

\subsection{Classical dispersive frame} 
\label{sub:classical_dispersive_frame}
We now focus on the qubit Hamiltonian~\eqref{eqn:H_0_prime}. Since $\proj{\alpha}$ has a time dependence involving the drive frequencies, this Hamiltonian is that of a qubit driven with multiple direct drives. We have not yet computed the amplitude of the fields yet and we will do so now taking $\alpha = \sum_i \alpha_i =  \sum_{d,i} \alpha_{i,d} e^{-i\omega_d t} + \alpha_{i,s} e^{-i\omega_s t}$. This choice assumes that the multiple drives are spread out enough in frequency such that one drive does not contribute significantly to the field oscillating at another drive's frequency. We therefore take
\begin{equation}
	\begin{split}
		H_0' &\approx \proj{\omega} + \sum_{i=0}^{M-2} \sum_d g_i \alpha_{i,d} e^{-i\omega_d t} \proj{i+1,i} + \hc \\
		&\quad + \sum_{i=0}^{M-2} g_i\alpha_{i,s} e^{-i\omega_s t} \proj{i+1,i} + \hc.
	\end{split}
\end{equation}
Transitions $\ket{i}\leftrightarrow \ket{i+1}$ are then driven by an off-resonant drive with amplitude $g_i\alpha_{i,d}$ and frequency $\omega_d$, as well as by a quasi-resonant drive with amplitude $g_i\alpha_{i,s}$ and frequency $\omega_s$. Focussing for now on the drives $\omega_d \neq \omega_s$, the first line of this Hamiltonian can be approximately diagonalized with an analog of the dispersive transformation~\eqref{eqn:quantum_dispersive_transformation}
\begin{equation}
	\label{eqn:classical_dispersive_transformation}
	\tD_C = \exp\lsb \sum_{i=0}^{M-2} \xi_i^* \proj{i,i+1} - \xi_i \proj{i+1,i}\rsb,
\end{equation}
where $\xi_i$ is a classical analog of the operator $\lambda_i \ad$. Because of this analogy, we will refer to this as the classical dispersive transformation. This transformation is performed on the master equation~(\ref{eqn:master_equation_prime}) in Appendix~\ref{annsec:dispersive_transformations} where we take 
\begin{equation}
	\label{eqn:xi_i}
	\xi_i = \sum_d \xi_{i,d} e^{-i\omega_d t},
\end{equation}
with $\xi_{i,d} = \varlambda_i^d \alpha_{i,d}$. 

In the spirit of the dispersive transformation, $\tD_C$ assumes an off-resonant driving and therefore cannot be applied to transform the spectroscopy drive $s$. When doing the transformation, we drop time-dependent terms involving two different drive frequencies $\omega_{d_1} \pm \omega_{d_2}$ under the rotating wave approximation. We also assume that for the purpose of getting the qubit transition frequencies, $\alpha_{i,d} = \alpha_{0,d}$. This is the same as taking $|\alpha_{i} - \alpha_{i+1}|$ to be small. Essentially, we assume that the difference in the pointer states is not important to describe the value of the qubit transition frequencies, but is important to describe their widths. In other words, we say that the mean transition frequency depends on the mean cavity field, which is approximately $\alpha_{0,d}$ at low spectroscopy power (for the qubit is mostly in its ground state), while the width of the transition frequencies depend on the deviation of the cavity field from $\alpha_{0,d}$.

Performing the above transformation on the qubit Hamiltonian $H_0'$ to fourth order in perturbation theory together with the simplifications just outlined, we find
\begin{equation}
	\label{eqn:H_0_second}
	H_0'' = \sum_{i=0}^{M-1} \omega_i'' \proj{i,i} + \sum_{i=0}^{M-2} g_i\alpha_{i,s} e^{-i\omega_s t} \proj{i+1,i} + \hc,
\end{equation}
where
\begin{equation}
	\label{eqn:omega_second}
	\omega_i'' = \omega_i + \sum_{d} \varS_i^d |\alpha_d|^2 + \frac{1}{4} \sum_{d} \varK_i^d|\alpha_d|^4,
\end{equation}
are the ac-Stark shifted qubit frequencies and
\begin{equation}
	\label{eqn:classical_stark_shift_coefficients}
	\begin{split}
		\varS_i^d &\equiv -(\varchi_i^d - \varchi_{i-1}^d), \\
		\varK_i^d &\equiv -4\varS_i^d(|\varlambda_i^d|^2+|\varlambda_{i-1}^d|^2) \\
		&\quad - (3\varchi_{i+1}^d|\varlambda_i^d|^2-\varchi_i^d|\varlambda_{i+1}^d|^2) \\
		&\quad + (3\varchi_{i-2}^d|\varlambda_{i-1}^d|^2-\varchi_{i-1}^d|\varlambda_{i-2}^d|^2), 
	\end{split}
\end{equation}
are the quadratic and quartic ac-Stark shift coefficients with
\begin{equation}
	\label{eqn:varlambda_varchi}
	\begin{split}
		\varlambda_i^d &\equiv \frac{-g_i}{\omega_{i+1}-\omega_i-\omega_d}, \\
		\varchi_i^d &\equiv -g_i\varlambda_i^d = \frac{g_i^2}{\omega_{i+1}-\omega_i-\omega_d}.
	\end{split}
\end{equation}
We note that $g_i = 0$ for all $i \notin [0,M-2]$ in the initial model such that terms with a negative index or an index above $M-2$ on the right hand side of the equations above vanish. Comparing these expressions with equations (3a) and (3b) of Ref.~\cite{Boissonneault2010}, we highlight a few differences. First, both $\varS_i^d$ and $\varK_i^d$ now depend on the drive frequency $\omega_d$ instead of the resonator frequency $\omega_r$. As explained in section~\ref{sub:detuned_measurement_drive}, this follows from considering that the driving photons can be at a frequency significantly detuned from $\omega_r$. Actually, Eqs.~\pref{eqn:classical_stark_shift_coefficients} and \pref{eqn:varlambda_varchi} also hold for linear cQED, where the measurement drive is in practice chosen to be quasi-resonant with $\omega_r$. Next, the equation for $\varS_i^d$ does not involve terms of higher order than $g^2$. In Ref~\cite{Boissonneault2010}, these higher-order terms came from choosing a specific order for ladder operators when computing $K_i$ (i.e. $\ad\ad a a = \ad a\ad a + \ad a$). Here, the field is classical and there is no such ordering choice to be made. Finally, in Ref.~\cite{Boissonneault2010}, a second-order coupling caused by two-photon transitions was diagonalized, yielding fourth order corrections. This second-order coupling is however only significant in the straddling regime where the resonator frequency is between two qubit transition frequencies~\cite{Koch2007}. Since we are not considering this regime here, this two-photon transition was neglected. 

The next step is to apply the transformation $\tD_C$ on $H_1'$ to find
\begin{equation}
	\label{eqn:H_1_second}
	\begin{split}
		H_1'' &\approx \sum_{i=0}^{M-2} g_i(\ad+a)(\proj{i,i+1}+\proj{i+1,i}) \\
		&\quad + (G'' \ad + {G''}^\dag a) + H_{\rm SB},
	\end{split}
\end{equation}
where $G''$ can be found in \eqref{eqn:G_second}. The Hamiltonian $H_{\rm SB}$, whose definition can be found in \eqref{eqn:H_SB}, corresponds to red and blue sideband transitions. This Hamiltonian is the multi-level equivalent of the one obtained in Eq.~(B10) of Ref.~\cite{Blais2007} for a two-level system driven by two detuned drives and experimentally studied in Ref.~\cite{Wallraff2007}. The drive strength $G''$ can be set to zero with a proper choice of the fields $\alpha_i$, yielding an undriven resonator in this frame. Assuming that $|\omega_{d_1}-\omega_{d_2}|$ is sufficiently large to neglect time-dependent cross terms, choosing $G''=0$ implies
\begin{equation}
	\label{eqn:condition_alpha_d}
	\begin{split}
		0 &= (\omega_r-\omega_d-i\tfrac{\kappa}{2})\alpha_{i,d} + (K-i\kappa_{\rm NL})|\alpha_i|^2\alpha_{i,d} \\
		&\quad  + K'|\alpha_{i}|^4\alpha_{i,d} + \epsilon_d + \lp\varS_i^d+\frac{1}{3!}\varK_i^d|\alpha_i|^2\rp \alpha_{i,d},
	\end{split}
\end{equation}
for each qubit-detuned drive and
\begin{equation}
	\label{eqn:condition_alpha_s}
	\begin{split}
		0 &= (\omega_r-\omega_s-i\tfrac{\kappa}{2})\alpha_{i,s} + (K-i\kappa_{\rm NL})|\alpha_i|^2\alpha_{i,s} \\
		&\quad + K'|\alpha_i|^4\alpha_{i,s} + \epsilon_s,
	\end{split}
\end{equation}
for the spectroscopy drive. In writing these expressions, we have again assumed that, even though $\alpha_i\ne\alpha_j$, these amplitudes are close enough to replace one by the other in order to uncouple the equations for $i\ne j$.

We stress that because \eqref{eqn:condition_alpha_d} contains the qubit-state dependent cavity pull, the solutions $\alpha_{i,d}$ obtained here go beyond the linear response theory for the response of the field to a change of the qubit state. As explained briefly in section~\ref{sub:limits_to_the_validity_of_linear_response_in_circuit_qed} and as we will detail further later, a linear response theory would instead have solutions of the form $\alpha_i = \bar\alpha + f(\varS_i)$, where $\bar\alpha$ would be the solution of \eqref{eqn:condition_alpha_d} with $\varS_i^d=\varK_i^d=0$ and $f(\varS_i)$ would be some linear function of the cavity pull.

We note that the equations for two different drives $d_1\ne d_2$ are coupled through the total field $|\alpha_i|$. However, in the interest of reproducing the results of Ref.~\cite{Ong2011}, from this point on we will consider only a single qubit-detuned drive which we will label $d=p$ (the pump drive) in addition to the spectroscopy drive $s$. This implies that only the first line of $H_1''$ will remain. For the purpose of calculating $\alpha_s$, we will also assume that the spectroscopy amplitude $\epsilon_s$ is small enough so that $|\alpha_{i,p}|\gg|\alpha_{i,s}|$ and that $\alpha_{i,p}\sim\alpha_{j,p}$, such that we can replace $\alpha_i$ by $\alpha_{i,p}\approx\alpha_p$ in the equation for $\alpha_{i,s}$. Finally, since in practice $K'\ll K\ll g$, performing the classical dispersive transformation on $H_2'$ would yield corrections smaller than those that we have kept so far. We therefore neglect those and take $H_2''=H_2'$.

Finally, applying the transformation $\tD_C$ on the dissipation yields the master equation in this doubly transformed frame
\begin{equation}
	\label{eqn:master_equation_second}
	\begin{split}
		\dot\rho'' &= -i\comm{H_s''}{\rho''} + \kappa''\sD[a]\rho'' + 2\gamma_\varphi\sD\lsb \proj{\varepsilon}\rsb\rho'' \\
		&\quad + \sum_{i=0}^{M-2} \gamma_{\downarrow,i}'' \sD\lsb \proj{i,i+1} \rsb\rho'' + \sum_{i=0}^{M-2} \gamma_{\uparrow,i}'' \sD\lsb \proj{i+1,i} \rsb\rho'' \\
		&\quad + \kappa\sD\lsb \proj{\alpha}\rsb\rho'' + \kappa_{\rm NL}\sD\lsb \proj{\alpha}^2\rsb\rho'' \\
		&\quad + \gamma \sD\lsb \sum_{i=0}^{M-1} \frac{g_i \xi_i - g_{i-1}\xi_{i-1}}{g_0} \proj{i,i}\rsb\rho'',
	\end{split}
\end{equation}
where
\begin{subequations}
	\label{eqn:rates_second}
	\begin{align}
		\label{eqn:kappa_second}
		\kappa'' &= \kappa + 4\kappa_{\rm NL}|\alpha_p|^2, \\
		\label{eqn:gamma_down_second}
		\gamma_{\downarrow,i}'' &\equiv \gamma \lp\frac{g_i}{g_0}\rp^2 + \gamma_{{\rm DD},i}'', \\
		\label{eqn:gamma_up_second}
		\gamma_{\uparrow,i}'' &\equiv \gamma_{{\rm DD},i}'', \\
		\gamma_{{\rm DD},i}'' &\equiv \lsb2\gamma_\varphi |\varepsilon_{i+1}-\varepsilon_i|^2 + \kappa|\alpha_{i+1,p}-\alpha_{i,p}|^2\rsb {\varlambda_i^p}^2 |\alpha_{i,p}|^2,
	\end{align}
\end{subequations}
and
\begin{equation}
	H_s'' = H_0'' + \sum_{i=0}^{M-2} g_i(\ad\proj{i,i+1}+a\proj{i+1,i}) + H_2''.
\end{equation}

These two transformations result in an ac-Stark shifted qubit that is driven with a spectroscopy drive of amplitude $\alpha_{s,i}$ and frequency $\omega_s$, coupled with a Jaynes-Cummings coupling to an undriven resonator whose frequency is shifted by the nonlinearity. This resonator sees additional relaxation $\kappa''>\kappa$ due to the two-photon-loss relaxation channel. The qubit sees its intrinsic dephasing at a rate $\gamma_\varphi$, as well as relaxation at rate $\gamma_{\downarrow,i}''$ and heating at rate $\gamma_{\uparrow,i}''$. These relaxation and heating rates are modified by dressed-dephasing~\cite{Boissonneault2008} (first term of $\gamma_{{\rm DD},i}$), but also dressed measurement-induced dephasing (second term). These rates were obtained assuming white noise for all the dissipation channels. If the noise is not white, the rate $\gamma_{{\rm DD},i}$ will depend on the noise spectra of the qubit dephasing and resonator relaxation channels at $\pm(\omega_{i+1,i}-\omega_d)$~\cite{Boissonneault2009}. In addition to intrinsic dephasing, the last two lines of \eqref{eqn:master_equation_second} contain three other sources of dephasing. The first term will yield measurement-induced dephasing~\cite{Gambetta2006}, while the second and the third represent respectively measurement-induced dephasing through the resonator two-photon loss decay channel and through the emission of an excitation by the qubit in its environment. While not measurable, this excitation in principle carries information about the qubit state and thus causes dephasing.

\subsection{Quantum dispersive frame and reduced master equation} 
\label{sub:quantum_dispersive_frame_and_reduced_master_equation}
The final effect that we would like our model to capture is the Lamb shift of the qubit frequencies due to vacuum fluctuations of the resonator. To obtain this shift, we perform the dispersive transformation $\tD$ of~\eqref{eqn:quantum_dispersive_transformation} on the master equation \eqref{eqn:master_equation_second}. Doing this while neglecting the photon population that is almost zero in the polaron frame [see discussion below \eqref{eqn:master_equation_prime}] yields the same master equation, but with the transformed Hamiltonian $H_s'''=H_0'''+H_2'''$ with
\begin{subequations}
	\begin{align}
		\label{eqn:H_0_third}
		H_0''' &= \proj{\omega'''} + \sum_{i=0}^{M-2} g_i\alpha_{i,s} e^{-i\omega_s t} \proj{i+1,i} + \hc, \\
		\label{eqn:H_2_third}
		H_2''' &= \lsb\omega_r'(\alpha) + \proj{S(\alpha)}\rsb\ada + \Upsilon{\ad}^2 + \Upsilon^*a^2
	\end{align}
\end{subequations}
and where the relaxation rate
\begin{equation}
	\gamma_{\downarrow,i}''' = \gamma_{\downarrow,i}'' + \lambda_i^2(\alpha) \kappa''
\end{equation}
has an added Purcell relaxation rate. The new Lamb-shifted frequencies $\omega_i{'''}$ are given by
\begin{subequations}
	\begin{align}
		\label{eqn:omega_i_third}
		\omega_{i}'''(\alpha) &\equiv \omega_i''(\alpha) + L_i(\alpha), \\
		L_i(\alpha) &\equiv \chi_{i-1}(\alpha), \\
		S_i(\alpha) &\equiv -(\chi_i(\alpha) - \chi_{i-1}(\alpha)), \\
		\chi_i(\alpha) &\equiv -g_i\lambda_i(\alpha), \\
		\lambda_i(\alpha) &\equiv \frac{-g_i}{\omega_{i+1}''(\alpha) - \omega_i''(\alpha) - \omega_r'(\alpha)}.
	\end{align}
\end{subequations}
Since the resonator and qubit frequencies are pulled by the classical field due respectively to the nonlinearity and the ac-Stark shift, the Lamb shift depends on these pulled frequencies, and therefore on the amplitude of the cavity field.

Finally, projecting the qubit onto its $\{\ket{0},\ket{1}\}$ subspace and tracing out the resonator degrees of freedom yields a reduced qubit master equation
\begin{equation}
	\label{eqn:master_equation_reduced}
	\dot\rho_q = -i\comm{H}{\rho_q} + \gamma_{\downarrow,0}'''\sD[\sigma_-]\rho_q + \gamma_{\uparrow,0}'''\sD[\sigma_+]\rho_q + \frac{\gamma_\varphi'''}{2}\sD[\sigma_z]\rho_q.
\end{equation}
In this expression, we have defined 
\begin{equation}
	\label{eqn:H_reduced}
	H = \frac{\omega_{10}'''}{2} \sigma_z + g_0(\alpha_{0,s} e^{-i\omega_s t}\sigma_- +\hc),
\end{equation}
 where $\omega_{10}''' \equiv \omega_1'''(\alpha)-\omega_0'''(\alpha)$, and
\begin{equation}
	\label{eqn:gamma_varphi_triple}
	\gamma_\varphi''' = \gamma_\varphi + \Gamma_{\varphi m},
\end{equation}
where
\begin{equation}
	\label{eqn:gamma_m}
	\begin{split}
		\Gamma_{\varphi m} &\equiv \frac{\kappa D^2}{2} + \frac{\kappa_{\rm NL}|\alpha_1^2-\alpha_0^2|^2}{2} \\
		&\quad + \ssum_d \frac{\gamma |2\varchi_0^d\alpha_{0,d}-\varchi_1^d\alpha_{1,d}|^2}{2g_0^2},
	\end{split}
\end{equation}
and $D\equiv|\alpha_1-\alpha_0|$ is the distance between the pointer states. In the equation for the effective dephasing rate $\Gamma_{\varphi m}$, we see the measurement-induced dephasings due to single-photon cavity losses (second term), to two-photon cavity losses (third term), and to the information carried out by the excitation emitted when the qubit relaxes. While these three channels leak information about the qubit state, only the single-photon cavity loss channel is usually monitored. Moreover, since in practice $\kappa \gg \gamma\varchi^2/g_0^2,\kappa_{\rm NL}$, only this last channel will convey any significant amount of information and contribute to qubit dephasing.

\section{Backaction on the qubit} 
\label{sec:backaction_on_the_qubit}
Following the reduced qubit model derived in section~\ref{sec:a_reduced_qubit_model}, here we revisit the results presented in section~\ref{sec:linear_circuit_qed_in_a_nutshell} for the dispersive regime of linear circuit QED. In this section, we compare the theoretical model to experimental data and numerical simulations. The parameters used throughout are given in the caption of Fig.~\ref{fig:spectrum}. These parameters were adjusted to fit independent spectroscopic and time domain measurements of the device used in Ref.~\cite{Ong2011}. This device was composed of a transmon qubit~\cite{Koch2007} coupled to a coplanar waveguide resonator made nonlinear by a Josephson junction embedded in its central conductor. 

In subsection~\ref{sub:experiment_and_qubit_spectra}, we first quickly present the experiment already described in Ref.~\cite{Ong2011}. We then look more precisely at the Lamb and ac-Stark shifts of the qubit transition frequency $\omega_{1,0}$ in subsection~\ref{sub:ac_stark_and_lamb_shift} and at its linewidth in subsection~\ref{sub:measurement_induced_dephasing}. 

\subsection{Experiment and qubit spectra} 
\label{sub:experiment_and_qubit_spectra}
In Ref.~\cite{Ong2011}, we presented spectroscopic measurements of a transmon qubit coupled to a driven nonlinear resonator. The qubit was probed through the resonator with a drive of amplitude $\epsilon_s$ and frequency $\omega_s\sim\omega_{1,0}$. Meanwhile, that  resonator was pumped with a drive of amplitude $\epsilon_p$ and frequency $\omega_p\sim\omega_r$. The pump field was applied long before the qubit probe was turned on, enabling the resonator to reach its stationary state. Two detunings between the 
pump frequency $\omega_p$ and the resonator frequency $\omega_r$ where studied in detail. This was done in order to explore both the parametric amplification and the bifurcation regimes. Consequently, two biasing points $\omega_p/2\pi=(6430,6450)$~MHz corresponding to $\Omega/\Omega_C=(3.1,0.7)$, are presented below. Here, we redefined $\Omega$ with respect to the effective resonator frequency as pulled by the qubit in the ground state rather than the bare resonator frequency. These two biasing points are illustrated by the two vertical lines in the stability diagram of Fig.~\ref{fig:bifurcation_phase_diagram} (b). After probing the qubit, a bifurcation measurement was performed in order to determine the probability $P(\ket{1})$ that the qubit was excited by the probe drive. 

\begin{figure}
	\centering
	\includegraphics[width=0.95\hsize]{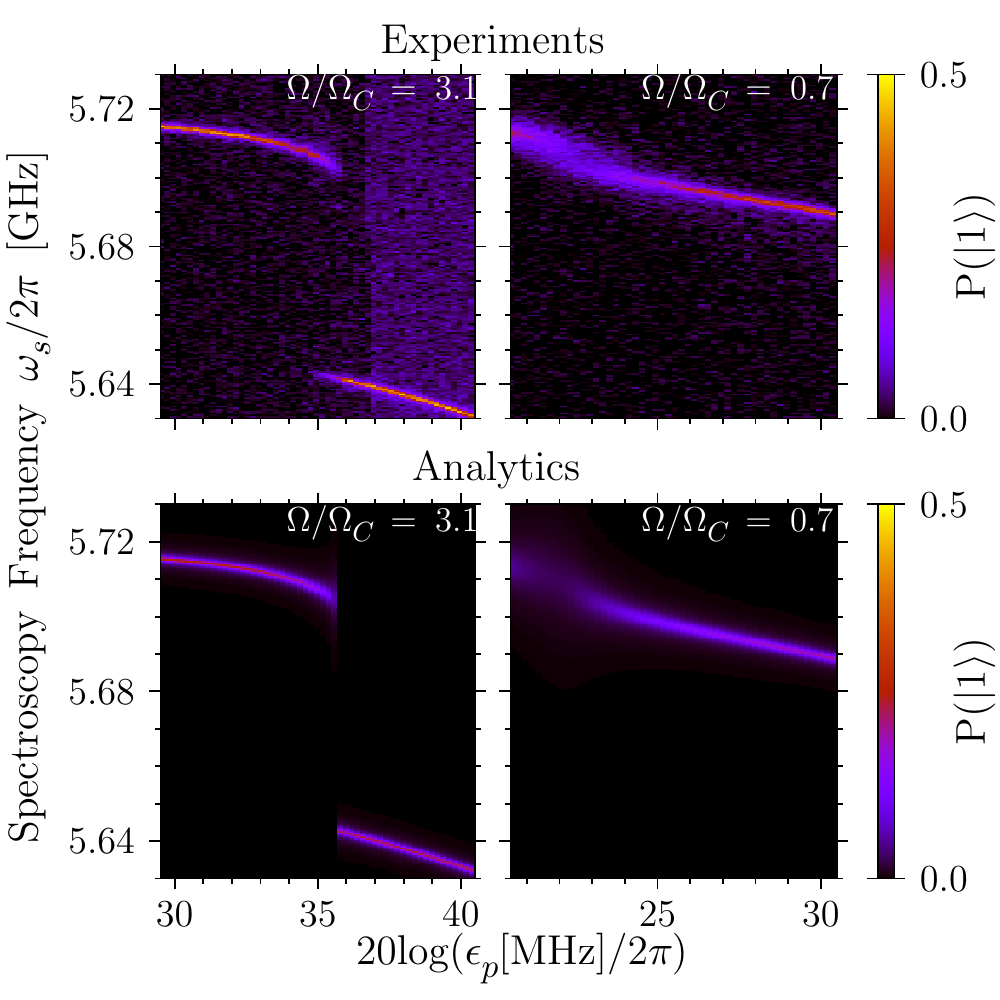}
	\caption{(Color online) Experimental (c.f. Ref.~\cite{Ong2011}) and analytical qubit excited state $\ket{1}$ population for the two operating points indicated in Fig.~\ref{fig:bifurcation_phase_diagram}. The qubit is a transmon with bare parameters $(\omega_{1,0},\omega_{2,1},\gamma,\gamma_\varphi)/2\pi = (5720,5421.6,0.22,0.25)$~MHz. The resonator's bare parameters are $(\omega_r,K,K',\kappa,\kappa_{\rm NL})/2\pi=(6453.5,-0.625,-0.00125,9.6,0)$~MHz and the qubit-resonator couplings are $(g_0,g_1)/2\pi=(42.4,58.4)$~MHz. These parameters were chosen to fit those of Ref.~\cite{Ong2011}. Couplings to higher transitions as well as higher transition frequencies can be computed from the transmon Hamiltonian~\cite{Koch2007}. The experimental attenuation required to link the experimental power in dB to the theoretical parameter $\epsilon_p$ was calibrated in Ref.~\cite{Ong2011}. Top: experimental spectroscopy results from Ref.~\cite{Ong2011}. Bottom: analytical stationary solution \eqref{eqn:P_ket_1}. Left (Right) : pump frequency $\omega_p/2\pi=6430(6450)$~MHz, corresponding to $\Omega/\Omega_C=3.1(0.7)$. These points are identified by vertical lines on Fig.~\ref{fig:bifurcation_phase_diagram}. The amplitude $\epsilon_s/2\pi=3$~MHz was chosen to fit the experimental power broadening of the qubit lines. }
	\label{fig:spectrum}
\end{figure}

The resulting experimental spectra are presented in the top panels of Fig.~\ref{fig:spectrum} for $\Omega/\Omega_C=3.1$ (top left) and $\Omega/\Omega_C=0.7$ (top right) as a function of the pump drive amplitude. The pump amplitude (horizontal axis) is converted to a logarithmic scale to match the experimental power in decibels, up to a constant offset that was calibrated in Ref.~\cite{Ong2011}. In the bifurcation regime (top left, $\Omega/\Omega_C=3.1$), we clearly see the jump in the qubit frequency associated with the jump from the low amplitude to the high amplitude dynamical states of the resonator. We also see that the line remains narrow and actually tends to narrow down at higher powers. In the parametric amplification regime (top right), we see a more monotonous shift of the qubit line with the measurement power with an important broadening around $20\log(\epsilon_p/2\pi)=22$. 

These spectra are then compared to the analytical steady-state solution of the reduced qubit master equation~\eqref{eqn:master_equation_reduced} in the bottom panels. The exact analytical solution of this equation yields~\footnote{We note that for Figs.~\ref{fig:spectrum}, \ref{fig:ac_stark}, and \ref{fig:measurement_induced_dephasing}, we set the dressed-dephasing rate $\gamma_{{\rm DD},i}''=0$ in the analytical model. This is done because we have derived the reduced model assuming white noise, whereas it is known that the rates can greatly depend on the noise spectrum~\cite{Boissonneault2009}. Considering white noise and not assuming $\gamma_{{\rm DD},i}''=0$ would result in a difference of background population at the threshold of bifurcation, which is not observed experimentally.}
\begin{equation}
	\label{eqn:P_ket_1}
	P(\ket{1}) = \frac{\gamma_{\uparrow,0}'''(\gamma_2^2+\delta^2) + 2\gamma_2|g_0\alpha_{0,s}|^2}{(\gamma_{\uparrow,0}'''+\gamma_{\downarrow,0}''')(\gamma_2^2+\delta^2) + 4\gamma_2|g_0\alpha_{0,s}|^2},
\end{equation}
where 
\begin{equation}
	\label{eqn:gamma_2}
	\gamma_2 \equiv \gamma_\varphi''' + \frac{\gamma_{\downarrow,0}'''+\gamma_{\uparrow,0}'''}{2},
\end{equation}
and $\delta\equiv \omega_{1,0}'''-\omega_s$.

When comparing the experimental to the analytical spectra, we notice small deviations between the background level as well as the amplitude of the spectroscopy lines. Aside from the limits of our model, three effects can cause these deviations. First, there is experimental thermal noise --- which should not exceed $50~$mK --- that is not taken into account in the theory and may yield a minor thermal qubit excited state population. Second, the experimental excited state population is extracted from the probability of bifurcation, which can yield an error of at most 0.05 in the estimated population. Third, the correspondance between the theoretical amplitude $\epsilon_s$ of the spectroscopy drive and the experimental amplitude could not be calibrated as precisely as the calibration provided by the ac-Stark shift for the pump drive~\cite{Ong2011}. Overlooking these deviations, other experimental features such as the spectroscopy lines' position and width are qualitatively reproduced by our analytical spectrum. In the following sections, we quantitatively compare these to our model. 

\subsection{Lamb and ac-Stark shifted qubit frequency} 
\label{sub:ac_stark_and_lamb_shift}
\begin{figure}
	\centering
	\includegraphics[width=0.95\hsize]{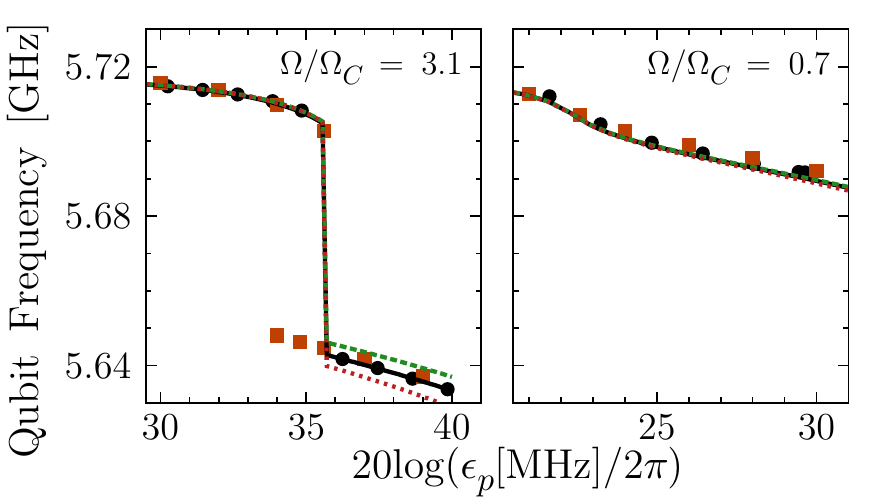}
	\caption{(Color online) Lamb and ac-Stark shifted qubit frequency as a function of the drive strength $\epsilon_p$ for the two operating points indicated in Fig.~\ref{fig:bifurcation_phase_diagram} and used in Ref.~\cite{Ong2011}. Parameters are the same as Fig~\ref{fig:spectrum}. Points are experimentally (black circles) and numerically (orange squares) extracted qubit transition frequency $\omega_{1,0}$ . Lines are analytically computed $\omega_{1}'''-\omega_0'''$ with the complete \eqref{eqn:omega_i_third} (full black lines), when setting $\varK_i^p=0$ (dotted red lines) or when taking $\omega_p=\omega_r$ in $\varS_i^p$ and $\varK_i^p$ (dashed green lines). }
	\label{fig:ac_stark}
\end{figure}
The experimental spectra presented in Fig.~\ref{fig:spectrum} were fitted using Lorentzian and the peak positions and widths were extracted from those fits, yielding the qubit transition frequency and dephasing rate. We also numerically integrated the multi-level Jaynes-Cummings master equation (\ref{eqn:master_equation}) to obtain numerical spectra that were fitted using the same procedure. The qubit frequency extracted from experimental (black circles) and numerical (orange squares) spectra is plotted in Fig.~\ref{fig:ac_stark} as a function of the pump power for the two operating points. Numerical simulations and experimental data almost coincide, suggesting that the initial master equation~(\ref{eqn:master_equation}) contains all the relevant physics. 

We then compare these data points to three versions of the dispersive approximation. Full black lines correspond to the complete equation~(\ref{eqn:omega_i_third}), dotted red lines correspond to the second order approximation for the dispersive shift (i.e. $\varK_i^p=0$), and dashed green lines correspond to setting $\omega_p=\omega_r$ when calculating $\varS_i^p$ and $\varK_i^p$. Since the parametric amplification regime (right panel) correspond to a pump drive very slightly detuned from the resonator frequency, as well as to a low number of photon ($n\sim20$), all three curves almost coincide in this regime. 

On the other hand, in the bifurcation regime (left panel), both the pump-resonator detuning and the number of photons after bifurcation are larger ($n\sim50$), yielding a significant difference between the three curves above bifurcation. We see that the assumption $\omega_p=\omega_r$ (dashed green lines), which as discussed in Sec.~\ref{sub:detuned_measurement_drive} is often made when calculating the ac-Stark shifts, yields a shift that is too small. This is expected since assuming $\omega_p=\omega_r$ yields a larger qubit-pump detuning, and correspondingly smaller values of $\varS_i^p$ and $\varK_i^p$. This effect can also be confirmed at lower power although it is not visible in these plots. We also see that the second order approximation (dotted red lines) yields a dispersive shift that is too large. This is also expected since the sign of each order in perturbation theory alternates sign in the dispersive regime and since  the fourth order is contained in the full model. 

With this model, the qubit can be used as a tool to characterize the nonlinear resonator. Indeed, the distance between the resonator's low and high amplitude states at the threshold of bifurcation directly depends on the resonator nonlinearity $K$ and the drive frequency $\omega_p$ and amplitude $\epsilon_p$. While experimentally $\omega_p$ is known to a very high precision, the resonator nonlinearity $K$ can only be estimated to about $\pm30\%$ from the design parameters due to its nonlinear dependence on sample parameters~\cite{Ong2011}. Moreover, the experimental line attenuation $A$ between the source and the input of the sample --- which is required to make the correspondance between the experimental power $P_p$ and the theoretical parameter $\epsilon_p$ --- can only be estimated up to about $2$~dB~\cite{Ong2011}. Performing a series of spectroscopic measurements for many pump frequencies $\omega_p$ and fitting the extracted qubit frequencies to the model derived here then makes it possible to extract both $K$ and $A$ with improved precision. This was done in Ref.~\cite{Ong2011} and resulted in an uncertainty of $2.4\%$ for $K$ and $0.2~$dB for $A$; a ten-fold improvement in precision.

\subsection{Qubit linewidth and validity of linear response} 
\label{sub:measurement_induced_dephasing}
We now examine the linewidth of the qubit transition. We know that, in addition to the intrinsic dephasing rate $\gamma_{2,\rm int} = \gamma_\varphi + \gamma/2$, the lines are broadened by measurement-induced dephasing~\cite{Gambetta2006} and by dressed-dephasing~\cite{Boissonneault2008}. In addition, there is always some power broadening due to the finite spectroscopy power. Here, we are mostly interested in the measurement-induced dephasing and how it is modified by the nonlinear nature of the resonator. The experiments presented in Ref.~\cite{Ong2011} and whose results are reproduced here were therefore carried in a regime where  power broadening is small. Moreover, since there is no dependence of the experimental background population over the pump power, we assume that dressed-dephasing is also negligible due to a small amplitude of dephasing noise at GHz frequencies. The only additional dephasing source is therefore measurement-induced dephasing in $\gamma_\varphi'''$ given in \eqref{eqn:gamma_varphi_triple} and in practice is dominated by the $\kappa|\alpha_1-\alpha_0|^2/2$ contribution. 

\begin{figure}
	\centering
	\includegraphics[width=0.95\hsize]{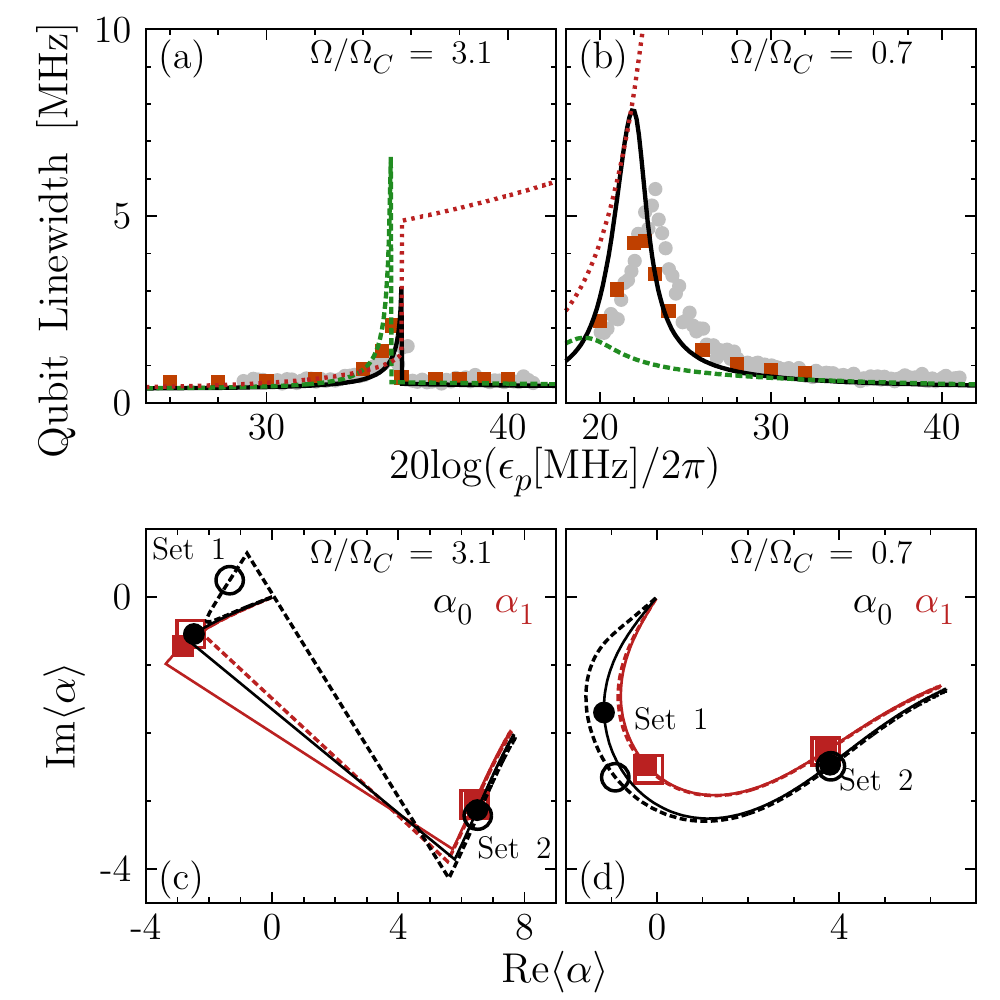}
	\caption{(Color online) (a) and (b) : Qubit line's half-width at half-maximum as a function of the drive strength $\epsilon_p$ for the two operating points indicated in Fig.~\ref{fig:bifurcation_phase_diagram} and used in Ref.~\cite{Ong2011}. Points are experimentally (grey circles) and numerically (orange squares) extracted qubit linewidths. Lines are analytical solutions corresponding to $\gamma_2$ where the fields $\alpha_i$ are computed according to the nonlinear response \eqref{eqn:condition_alpha_d} (full black lines), to linear response \eqref{eqn:linear_response_alphad} (dashed green lines) and by replacing $\kappa|\alpha_1-\alpha_0|^2/2$ with the linear resonator result \eqref{eqn:Gamma_m_Gambetta2006} (dotted red lines). (c) and (d) : Phase space representation of the fields $\alpha_0$ (black lines, circles) and $\alpha_1$ (red lines, squares) as given by the linear (dashed lines, empty symbols) and nonlinear (full lines, full symbols) response theories. All four symbols of a given set (1 or 2) correspond to a given pump amplitude $\epsilon_p$. }
	\label{fig:measurement_induced_dephasing}
\end{figure}

We present in Fig.~\ref{fig:measurement_induced_dephasing} the half-width at half-maximum of the spectroscopy lines as a function of the pump power for the two operating points $\Omega/\Omega_C=3.1(a),0.7(b)$. Grey circles (orange squares) are again the widths extracted from experimental (numerical) data. Full black lines are the analytical widths $\gamma_2/2\pi$ given by \eqref{eqn:gamma_2}. Dashed green lines are the same as the full black lines, but using linear response theory for the fields $\alpha_{i,p}$ instead of the solutions of \eqref{eqn:condition_alpha_d}. More precisely, we obtained the dashed green lines taking
\begin{equation}
	\label{eqn:linear_response_alphad}
	\alpha_{i,p} = \bar\alpha - \frac{\varS_i^p\bar\alpha}{(\omega_r-\omega_p - i\tfrac{\kappa}{2}) + 3K |\bar\alpha|^2},
\end{equation}
where $\bar\alpha$ is the solution of \eqref{eqn:condition_alpha_d} with $\varS_i^p = \varK_i^p = 0$. Finally, dotted red lines we obtained by replacing $\Gamma_{\varphi m}$ by the result of Ref.~\cite{Gambetta2006} for a linear resonator
\begin{equation}
	\label{eqn:Gamma_m_Gambetta2006}
	\Gamma_{m}^{\rm Linear} = \frac{\kappa}{2} \frac{2(|\alpha_{1,p}|^2+|\alpha_{0,p}|^2)\chi^2}{\kappa^2/4+\chi^2+(\omega_r-\omega_p)^2},
\end{equation}
where $\chi = \varS_1^p - \varS_0^p$.

The first striking observation is that, contrary to circuit QED with a linear resonator~\cite{Schuster2005}, the linewidth does not strictly increase with the drive power or equivalently with the number of photons in the resonator. In fact, in the bifurcation regime [Fig.~\ref{fig:measurement_induced_dephasing}(a)], the linewidth shows a sharp maximum at the bifurcation power, whereas in the parametric amplification regime, the linewidth shows a smooth maximum at a power that corresponds to the maximum gain of the amplifier~\cite{Ong2011}. This is illustrated by the lack of even qualitative agreement between both experimental and numerical data points and the result expected for a linear resonator (dotted red line). 

Narrowing  of the linewidth at high power is predicted both by the nonlinear (full dark lines) and the linear (dashed green lines) response theory. However, while both give a qualitative agreement with experimental and numerical data points, only the nonlinear response theory gives a quantitative one. In the bifurcation regime ($\Omega/\Omega_C=3.1$), the nonlinear response theory reproduces the experimental behavior with good accuracy on the whole range of powers, whereas linear response predicts bifurcation at too low power and linewidths twice as large at bifurcation. In the parametric amplification regime ($\Omega/\Omega_C=0.7$), only the nonlinear response solution gives semi-quantitative agreement near the maximum linewidth, while linear response theory predicts a much lower linewidth.  However, even the nonlinear response solution mispredicts the linewidth when it is above $\sim5$~MHz. We explain this by the breakdown of the $|\alpha_1-\alpha_0| < 1$ approximation, which corresponds to a measurement-induced dephasing rate of about $\Gamma_{\varphi m} \sim \kappa/4\pi\sim5$~MHz. 

To understand the non-monotonous behavior of the linewidth with drive power, we refer to Figs.~\ref{fig:measurement_induced_dephasing} (c) and (d), where we plot the value of the fields $\alpha_{0(1),p}$ as black (red) lines in the complex plane for the two operating points, for a range of power $\epsilon_p/2\pi \in [0,150]~$MHz and for nonlinear (full lines) and linear (dashed lines) response solutions. We see with these plots that even though the number of photons increases as the distance to the origin grows, the distance between the solutions $\alpha_{1,p}$ and $\alpha_{0,p}$ does not. In fact, the distance $D$ can be as small at higher power than at small power. 

For reference purposes, we also plot two sets of four points in pannels (c) and (d). Each set corresponds to a given pump amplitude $\epsilon_p$, for nonlinear (full symbols) and linear (empty symbols) theory, and for $\alpha_0$ (black circles) and $\alpha_1$ (red squares). Comparing the points within a given set of four points, we can see that a larger distance between a circle and its corresponding square --- and hence the larger the gain of the amplifier --- correspond to a larger disagreement between the linear and nonlinear solutions (distance between a full and a corresponding empty symbol).

\begin{figure}
	\centering
	\includegraphics[width=0.95\hsize]{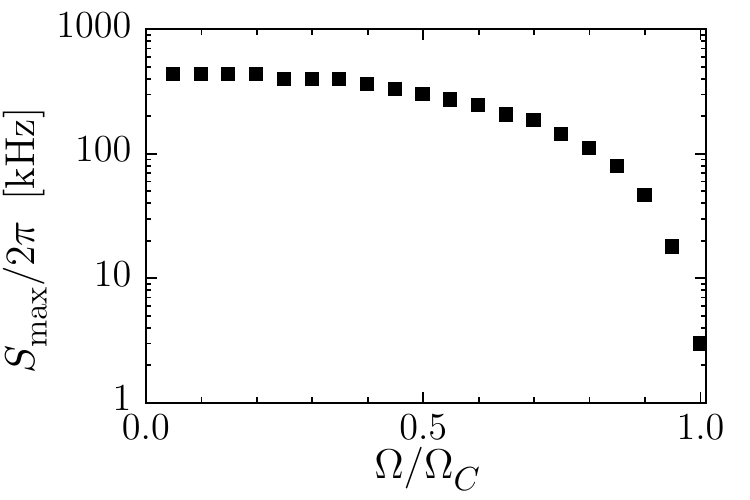}
	\caption{Maximum dispersive coupling $S_{\rm max}$ for which the linear response theory is valid with 10\% uncertainty at the region of maximum gain as a function of the reduced detuning $\Omega/\Omega_C$ in the parametric amplification regime.}
	\label{fig:validity_linear_response}
\end{figure}

We can compute a range of validity of the linear response theory by computing the fields $\alpha_{i,p}$ to second order (i.e.~quadratic response theory). If we define $\alpha_{i,p} = \bar\alpha + \alpha_{i,p}^{(1)} + \alpha_{i,p}^{(2)}$, where $\alpha_{i,p}^{(1)}$ is the second term of equation~\eqref{eqn:linear_response_alphad} and $\alpha_{i,p}^{(2)}$ is the next order correction, linear response theory will be valid if the ratio $r = \alpha_{i,p}^{(2)}/\alpha_{i,p}^{(1)}$ is small. Since for a qubit measurement, the signal that is amplified is a frequency shift $S = \pm (\varS_1-\varS_0)$, we can  define a maximal value of $S$ that allows $r$ to be smaller than a threshold $r_t$ in the region of highest gain. This maximal value $S_{\rm max}$, computed using a conservative value of 10\% for the ratio of the quadratic correction over the linear correction, is plotted in Fig.~\ref{fig:validity_linear_response} as a function of the reduced detuning $\Omega/\Omega_C$. We see that the maximal coupling for the parameters given in the caption of Fig.~\ref{fig:spectrum},  typical for circuit QED,  never exceeds about $0.5~$MHz. Moreover, the maximal coupling in fact vanishes when approaching the critical detuning $\Omega_C$. This maximal coupling is to be compared with the resonator linewidth $\kappa$ in order to determine if it is viable for a qubit measurement. With a realistic criteria of $\chi\ge 0.2\kappa$ to get a good measurement, one therefore needs either $\kappa/2\pi\sim1$~MHz or a smaller nonlinearity $K$ in order for linear response theory to be valid in this system. The former however implies a longer measurement time, while the latter implies a smaller gain, both impairing the efficiency of the measurement. It therefore seems unlikely that linear response theory will be sufficient to describe any superconducting qubit readout using a nonlinear resonator until the qubit lifetimes become long enough for longer measurement time to be viable. 

\subsection{Quantum limit to the added noise} 
\label{sub:quantum_limit_to_the_added_noise}
Using the results presented in Fig.~\ref{fig:measurement_induced_dephasing}, we can try to answer the question of whether or not a dispersive homodyne measurement using a nonlinear resonator can reach the quantum limit $\Gamma_{\varphi m} = \Gamma_{meas}/2$ as is the case for a linear resonator~\cite{Gambetta2008}. Indeed, assuming small squeezing, if one were to make a homodyne measurement using the pump drive, the measurement rate would be given by $\Gamma_{meas}=\kappa|\alpha_1-\alpha_0|^2$~\cite{Gambetta2008}. Since this measurement rate is exactly twice the dominant part of the measurement induced dephasing caused by these same pump photons $\Gamma_{\varphi m}$ given at~\eqref{eqn:gamma_m}, we can say that the quantum limit is reached if the theoretical prediction fits the experimental linewidth. If the experimental linewidth is larger than the theoretical prediction, it however means that the limit is missed. Finally, if the experimental linewidth is smaller than that predicted by the model, it means that one of the approximation is probably breaking down.

Looking at Fig.~\ref{fig:measurement_induced_dephasing}~(b), we then reach a different conclusion whether we consider linear or nonlinear response. Indeed, around $20\log(\epsilon_p/2\pi) \in [20,30]$, the experimental linewidth is much higher than the prediction from linear response, and we would therefore conclude that the quantum limit is missed by the measurement. This is qualitatively the same conclusion as the one obtained by Laflamme and Clerk~\cite{Laflamme2011}, also in a linear response theory. However, we know from Fig~\ref{fig:validity_linear_response}, that for  $\Omega/\Omega_C=0.7$ as in Fig.~\ref{fig:measurement_induced_dephasing}~(b), the maximum dispersive coupling supported by a linear response treatment is $S_\mathrm{max}/2\pi\sim200$~kHz, about four times smaller than the one used here. If we now compare the nonlinear response model prediction (black line), we see that it matches the experimental observations on a much wider range, and we recover the quantum limit in this range. There is also a regime where the theoretical prediction is above the experimental observation. This regime corresponds to a linewidth $\sim \kappa D^2/2 \gtrsim 5$~MHz since $\kappa/2\pi \sim 10$~MHz, and therefore to $D\gtrsim 1$, breaking the small distinguishability approximation that we have made. Therefore, while our result shows that the quantum limit can be reached with a nonlinear resonator, the question remains open in the case of large distinguishability or large squeezing where our model breaks down.

\section{Conclusion} 
\label{sec:conclusion}
In summary, we have derived an analytical model to describe the backaction of a driven nonlinear resonator on a multi-level qubit. This is done using unitary transformations, and especially using the polaron~\cite{Mahan2000,Irish2005,Gambetta2008} and  dispersive~\cite{Carbonaro1979,Boissonneault2009} transformations. We obtain a reduced model that contains the physics of the linear and quadratic ac-Stark shifts as well as the Lamb shift of the qubit frequencies. The model also contains dressed-dephasing~\cite{Boissonneault2009,Boissonneault2008,Wilson2010}, Purcell relaxation~\cite{Houck2008} and measurement-induced dephasing~\cite{Gambetta2008,Gambetta2006,Schuster2005}. Contrary to other theoretical models, both qualitative and quantitative agreements are found for the ac-Stark and Lamb shifted qubit transition frequencies as well as for the qubit linewidth.

Moreover, the model that we have derived here goes beyond some assumptions that are frequently made and that are valid in the case of a driven linear resonator, but not in the nonlinear case. These assumptions are the resonant driving of the resonator, the linear response of the resonator field to the qubit signal and the two-level character of the qubit. Considering detuned driving of the resonator yields linear and quadratic ac-Stark shifts that depend on the qubit-drive frequency detuning rather than the qubit-resonator frequency detuning and are therefore slightly different than usual dispersive shifts~\cite{Blais2004}. Going beyond linear response theory yields measurement-induced dephasing rates that are qualitatively different from those found with linear response and that are found to match the experimental and numerical data in most regimes considered. In particular, we show that the measurement-induced dephasing rate does not increase  with the measurement power or the number of photons, but rather with the distance between two pointer states $\alpha_1$ and $\alpha_0$ of the resonator fields. The precise quantitative agreement between the model and the experiment has also allowed us in Ref.~\cite{Ong2011} to characterize the nonlinearity of the resonator and the attenuation of the transmission line with an accuracy ten times better than what was otherwise achievable.

We have finally also shown that the results given by linear reponse theory are unlikely to apply to any high-fidelity qubit measurement using a nonlinear resonator. One consequence of this is to reopen the question of whether or not measurement with a nonlinear resonator is quantum limited in the amount of dephasing it causes on a qubit. Indeed, while Laflamme and Clerk~\cite{Laflamme2011} have shown that the quantum limit is missed by a factor $G$, the gain of the amplifier, this result was obtained in a linear response theory and therefore is not applicable in the systems considered here. This question then remains open and could be answered using a quantum trajectory approach as was done before for a linear resonator~\cite{Gambetta2008}.

\begin{acknowledgments}
We acknowledge discussions with M.~Dykman, A.~A.~Clerk, C.~Laflamme, J.~M.~Gambetta, D.~H.~Slichter, R.~Vijay and within the Quantronics group. We acknowledge support from NSERC, FQRNT, the Alfred P. Sloan Foundation, CIFAR, the ANR project Quantjo, the European project SCOPE and the Australian Research Council. We thank Calcul Qu\'ebec and Compute Canada for computational resources.
\end{acknowledgments}

\appendix
\section{Polaron transformation} 
\label{annsec:polaron_transformation}
In this appendix, we give the result of applying the polaron transformation~(\ref{eqn:polaron_transformation}) on the different parts of the master equation~(\ref{eqn:master_equation}). The building blocks from which all operators can be transformed are
\begin{subequations}
	\begin{align}
		a' &= a + \proj{\alpha}, \\
		\proj{i,i+1}' &= \proj{i,i+1} D^\dag(\alpha_i) D(\alpha_{i+1}) \approx \proj{i,i+1}, \\
		\proj{i,i}' &= \proj{i,i},
	\end{align}
\end{subequations}
where we noted the transformed-frame operator $O' \equiv \tP^\dag O \tP$. Using these relations, transforming the Hamiltonian $H_r$ yields
\begin{widetext}
	\begin{equation}
		\label{eqn:H_r_prime}
		\begin{split}
			H_r' &= \omega_r (\ada + \proj\alpha^* a + \proj\alpha \ad + |\proj\alpha|^2) + \frac{K}{2} \lsb |\proj\alpha|^4 + (2\ad|\proj\alpha|^2\proj\alpha + \hc) + 4\ada|\proj\alpha|^2 + ({\ad}^2\proj\alpha^2 + \hc)\rsb  \\
			&\quad  + \frac{K'}{3}\lsb |\proj\alpha|^6 + (3\ad|\proj\alpha|^4\proj\alpha + \hc) + 9\ada|\proj\alpha|^4 + (3{\ad}^2|\proj\alpha|^2\proj\alpha^2 + \hc)\rsb,
		\end{split}
	\end{equation}	
\end{widetext}
where we have dropped terms with more than two resonator ladder operators. This approximation assumes that $|\alpha| \gg |\mean{a}|$ in the transformed frame. We will see that with a proper choice of $\alpha$, the resonator in the transformed frame is close to its ground state. Transforming $H_q$ is trivial since it is diagonal in the qubit subspace and therefore commutes with the transformation and $H_q'=H_q$. Transforming the interaction Hamiltonian $H_I$ yields
\begin{equation}
	\label{eqn:H_I_prime}
	\begin{split}
		H_I' &= \sum_{i=0}^{M-2} g_i\lsb\ad + a\rsb \lsb \proj{i,i+1} + \proj{i+1,i}\rsb \\
		&\quad + \sum_{i=0}^{M-2} g_i\lsb \proj\alpha^* \proj{i,i+1} + \proj{i+1,i}\proj\alpha\rsb.
	\end{split}
\end{equation}
In obtaining this equation, we assumed that $D(\alpha_{i+1}-\alpha_{i})\approx 1$ and made a RWA for the second line. Not doing the RWA would only yield a small Bloch-Siegert shift to the qubit transition frequencies~\cite{Bloch1940}. However, we choose not to do a RWA on the first line at this point. This will allow us to get the sidebands Hamiltonian for a MLS, equivalently to what was done in Ref.~\cite{Blais2007} for a TLS. Transforming the drive Hamiltonians $H_s$ and $H_d$ is also trivial and yields
\begin{subequations}
	\begin{align}
		\label{eqn:H_d_prime}
		H_d' &= \sum_d \epsilon_d e^{-i\omega d t}(\ad+\proj{\alpha}^*) + \hc, \\
		\label{eqn:H_s_prime}
		H_s' &= \epsilon_s e^{-i\omega_s t} (\ad+\proj{\alpha}^*) + \hc.
	\end{align}
\end{subequations}
Finally, since the transformation $\tP$ moves the system to a time-dependent frame, a Hamiltonian
\begin{subequations}
	\begin{align}
		\label{eqn:H_tP}
		H_\tP &\equiv i \dot\tP^\dag \tP = (-i\dproj\alpha\ad + i\dproj\alpha^*a) - \ImaginaryPart[\proj\alpha\dproj\alpha^*],
	\end{align}
\end{subequations}
must be added to the total Hamiltonian in the transformed frame. 

For the dissipation, we can show that
\begin{equation}
	\begin{split}
		\sD[a']\rho' &= \sD[a]\rho' -i\comm{\frac{i}{2}\proj{\alpha}^*a + \hc}{\rho'}, \\
		&\quad + a\comm{\rho'}{\proj{\alpha}^*} + \hc.
	\end{split}
\end{equation}
In this equation, the second term is of Hamiltonian form and will be added to the Hamiltonian in the transformed frame. It is worth noting that, if $\rho'$ is the ground state of the resonator in this frame, the last line is equal to zero. For the two-photon dissipation, we get
\begin{equation}
	\begin{split}
		\sD[a'a']\rho' &\approx \sD[\proj{\alpha}^2]\rho' + 4\sD[a\proj\alpha]\rho' \\
		&\quad - i \comm{i\lp a|\proj\alpha|^2\proj\alpha^*+\frac12 {\proj{\alpha}^*}^2a^2\rp+\hc}{\rho'} \\
		&\quad + \lp 2a\proj\alpha + a^2\rp \comm{\rho'}{\proj\alpha^{*2}} + \hc,
	\end{split}
\end{equation}
where again, the last line is zero if $\rho'$ is the ground state of the system, and the second line is of Hamiltonian form and will be included in the Hamiltonian in this transformed frame. When obtaining this result, we assumed again that $|\alpha|\gg|\mean{a}|$, and neglected any term with more than two ladder operators. 

Since the polaron transformation commutes with $\proj{\varepsilon}$, $\sD[\proj{\varepsilon}]$ stays the same in the transformed frame. Moreover, since we assumed that $|\alpha_{i+1}-\alpha_{i}|<1$, we do not transform the dissipators $\sD[\proj{i,i+1}]$. 

\section{Dispersive transformations} 
\label{annsec:dispersive_transformations}
In this appendix, we transform the different parts of the master equation~(\ref{eqn:master_equation_prime}) according to the classical dispersive transformation $\tD_C$ given at \eqref{eqn:classical_dispersive_transformation}. We note $O'' \equiv \tD_C^\dag O'\tD_C$. To first order in $\xi$, we get
\begin{widetext}
	\begin{equation}
		\label{eqn:H_0_second_first_order}
			H_0'' \approx \frac{1}{0!}\proj\omega + \frac{1}{1!}\sum_{i=0}^{M-2} \omega_{i+1,i} (\xi_i^*\proj{i,i+1} + \xi_i\proj{i+1,i}) + \frac{1}{0!}\sum_{i=0}^{M-2} g_i (\alpha_i^* \proj{i,i+1}+\alpha_i\proj{i+1,i}) + H_{\tD_C},
	\end{equation}	
\end{widetext}
where
\begin{equation}
	H_{\tD_C} \equiv i\dot\tD_C \tD_C^\dag \approx \frac{1}{1!} \sum_{i=0}^{M-2} i\dot\xi_i^*\proj{i,i+1}-i\dot\xi_i\proj{i+1,i}.
\end{equation}
Assuming the form of \eqref{eqn:xi_i} for $\xi_i(t)$, we can compute $\dot\xi_i(t)$. Doing this and taking
\begin{equation}
	\ssum_{d} \lsb (\omega_{i+1,i}-\omega_d)\xi_{i,d} + g_i\alpha_{i,d}\rsb e^{-i\omega_d t} = 0,
\end{equation}
or equivalently
\begin{equation}
	\xi_{i,d} = \frac{-g_i\alpha_{i,d}}{\omega_{i+1,i}-\omega_d},
\end{equation}
makes the non-diagonal terms in \eqref{eqn:H_0_second_first_order} vanish. With this choice, transforming $H_0'$ to fourth order in perturbation theory and assuming large frequency differences and sums $|\omega_{d_1}\pm\omega_{d_2}|$ as well as small $|\alpha_{i,d}-\alpha_{i+1,d}|$ yields the Hamiltonian $H_0'' = \tD_C^\dag H_0'\tD_C + H_{\tD_C}$ given by \eqref{eqn:H_0_second}. 

In order to transform $H_1'$, we need to know how to transform a diagonal operator
\begin{equation}
	\tD_C^\dag\proj{x}\tD_C \approx \proj{x} + \sum_{i=0}^{M-2} (x_{i+1}-x_i)(\xi_i\proj{i+1,i} + \xi_i^*\proj{i,i+1}),
\end{equation}
and the off-diagonal operator $\Sigma_- \equiv \sum_{i=0}^{M-2} g_i \proj{i,i+1}$
\begin{widetext}
	\begin{equation}
		\begin{split}
			\tD_C^\dag \Sigma_- \tD_C &\approx \sum_{i=0}^{M-2} g_i\proj{i,i+1} + \sum_d \proj{\varS^d}\proj{\alpha_d} e^{-i\omega_d t} + \frac{1}{3!}\sum_d \proj{\varK^d} |\proj{\alpha_d}|^2\proj{\alpha_d} e^{-i\omega_d t} \\
			&\quad + \frac{1}{2!} \sum_{i=0}^{M-2} \sum_{d_1,d_2} \varlambda_i^{d_1}\lsb \varS_{i+1}^{d_2} - \varS_{i}^{d_2}\rsb\lsb\alpha_{d_1}\alpha_{d_2} e^{-i(\omega_{d_1}+\omega_{d_2})t}\proj{i+1,i}+\mathrm{h.c.}\rsb \\
			&\quad + \frac{1}{2!} \sum_{i=0}^{M-2} \sum_{d_1,d_2} \lsb 2\varlambda_i^{d_1}\lp-\varchi_{i+1}^{d_2} + \varchi_i^{d_2} - \varchi_{i-1}^{d_2}\rp-g_i\lp\varlambda_{i-1}^{d_1}\varlambda_{i-1}^{d_2}+\varlambda_{i+1}^{d_1}\varlambda_{i+1}^{d_2}\rp\rsb \lsb\alpha_{d_1}^*\alpha_{d_2} e^{i(\omega_{d_1}-\omega_{d_2})t}\proj{i,i+1}+\mathrm{h.c.}\rsb,
		\end{split}
	\end{equation}	
where we made the same assumptions as previously and $\varchi_i^d$, $\varlambda_i^d$, $\varS_i^d$ and $\varK_i^d$ are defined at equations~(\ref{eqn:classical_stark_shift_coefficients})-(\ref{eqn:varlambda_varchi}). These two equations can be combined and used to transform $H_1'$ yielding the result of \eqref{eqn:H_1_second} with
\begin{equation}
	\label{eqn:G_second}
	G'' = G' + \sum_d \lp\proj{\varS^d} + \frac{1}{3!} \proj{\varK^d}|\proj{\alpha}|^2\rp\proj{\alpha_d} e^{-i\omega_d t},
\end{equation}
and
	\begin{equation}
		\label{eqn:H_SB}
		\begin{split}
			H_{\rm SB} &\approx \sum_{i=0}^{M-2} \sum_{d_1,d_2} \lcb \varlambda_i^{d_1}\lsb \varS_{i+1}^{d_2} - \varS_i^{d_2}\rsb\lsb\alpha_{d_1}\alpha_{d_2} e^{-i(\omega_{d_1}+\omega_{d_2})t}\proj{i+1,i}+\mathrm{h.c.}\rsb\rcb\frac{\ad}{2!} \\
			&\quad + \sum_{i=0}^{M-2} \sum_{d_1,d_2} \lcb \lsb 2\varlambda_i^{d_1}\lp-\varchi_{i+1}^{d_2} + \varchi_i^{d_2} - \varchi_{i-1}^{d_2}\rp-g_i\lp\varlambda_{i-1}^{d_1}\varlambda_{i-1}^{d_2}+\varlambda_{i+1}^{d_1}\varlambda_{i+1}^{d_2}\rp\rsb\lsb\alpha_{d_1}^*\alpha_{d_2} e^{i(\omega_{d_1}-\omega_{d_2})t}\proj{i,i+1}+\mathrm{h.c.}\rsb\rcb \frac{\ad}{2!}, \\
			&\quad - \sum_{i=0}^{M-2}\sum_{d_1,d_2} \varlambda_{i}^{d_1}[\varS_{i+1}^{d_2}-\varS_i^{d_2}] \lsb \alpha_{d_1}\alpha_{d_2}  e^{-i(\omega_{d_1}+\omega_{d_2})t}\proj{i+1,i} + \alpha_{d_1}^*\alpha_{d_2} e^{i(\omega_{d_1}-\omega_{d_2})t}\proj{i,i+1}\rsb \ad \\
			&\qquad\qquad + \mathrm{h.c.}.	
		\end{split}
	\end{equation}
In this last Hamiltonian, the choice of the polaron frame~\eqref{eqn:condition_alpha_d} has already been made. Finally, transforming dissipators according to $\tD_C$ yields
	\begin{subequations}
		\begin{align}
			2\gamma_\varphi \sD\lsb \frac{\proj\varepsilon}{\varepsilon_1}\rsb \rho' &\rightarrow 2\gamma_\varphi \sD\lsb \frac{\proj\varepsilon}{\varepsilon_1}\rsb \rho'' + 2\gamma_\varphi\sum_{i=0}^{M-2}\ssum_{d} \frac{|\varepsilon_{i+1}-\varepsilon_{i}|^2}{\varepsilon_1^2} |\xi_{d,i}|^2(\sD[\proj{i+1,i}]\rho'' + \sD[\proj{i,i+1}]\rho'') \\
			\gamma\sum_{i=0}^{M-2}\lp\frac{g_i}{g_0}\rp^2 \sD[\proj{i,i+1}]\rho' &\rightarrow \gamma\sum_{i=0}^{M-2}\lp\frac{g_i}{g_0}\rp^2 \sD[\proj{i,i+1}]\rho'' + \gamma \sD\lsb \sum_{i=0}^{M-1} \frac{g_i \xi_i - g_{i-1}\xi_{i-1}}{g_0} \proj{i,i}\rsb\rho'' \\
			\kappa \sD\lsb \proj\alpha\rsb \rho' &\rightarrow \kappa \sD\lsb \proj\alpha\rsb \rho'' + \kappa\sum_{i=0}^{M-2}\ssum_{d} |\alpha_{i+1}-\alpha_{i}|^2|\xi_{d,i}|^2(\sD[\proj{i+1,i}]\rho'' + \sD[\proj{i,i+1}]\rho'').
		\end{align}
	\end{subequations}	
These transformed Hamiltonians and dissipators can be combined to obtain the master equation in the polaron and classical dispersive frame given in Eq.~(\ref{eqn:master_equation_second}).
\end{widetext}

\bibliography{/Users/mboisson/Documents/Articles/Biblio.bib}

\end{document}